\UseRawInputEncoding
\documentclass[prd,twocolumn,showpacs,superscriptaddress,nofootinbib]{revtex4-1}
\usepackage{amsfonts}
\usepackage{tikz}
\usepackage{pgfplots}
\usepackage{amsmath}
\usepackage{amssymb}
\usepackage{bm}
\usepackage{dcolumn}
\usepackage{graphicx}
\usepackage{graphics}
\usepackage{latexsym}
\usepackage{rotating}
\usepackage[colorlinks=true]{hyperref}
\usepackage[all]{hypcap} 
\usepackage{xspace} % Sensible space treatment at end of simple macros
\usepackage{xcolor}
\usepackage{mathrsfs}
\usepackage{siunitx}
\usepackage{multirow}
\usepackage{soul}
\usepackage{ulem}
\usepgfplotslibrary{fillbetween}
\definecolor {darkgreen}{rgb}{0.2,0.7,0.2}

\newcommand\be{\begin{equation}}
\newcommand\ee{\end{equation}}
\newcommand\bw{\begin{widetext}}
\newcommand\ew{\end{widetext}}

\newcommand{\bea}{\begin{eqnarray}}
\newcommand{\eea}{\end{eqnarray}}

\newcommand{\cm}{{\rm cm}}

\newcommand{\g}{{\rm g}}

\usepackage{cellspace} %
\newlength\figureheight
\newlength\figurewidth

\newlength\figureheightmed
\newlength\figurewidthmed

\newlength\figureheightlarge
\newlength\figurewidthlarge

\def\apj{Astrophys. J. \ }%
\def\apjl{Astrophys. J. Lett. \ }%
\def\aap{A\&A}%
\def\prc{Phys. Rev. C\ }%
\def\prd{Phys. Rev. D \ }%
\begin{document}
\title{The fate of twin stars on the unstable branch: implications for the formation of twin stars}

\author{Pedro L. Espino}
\affiliation{Department of Physics, University of Arizona, Tucson, AZ 
85721, USA}
\author{Vasileios Paschalidis}
\affiliation{Department of Physics, University of Arizona, Tucson, AZ 
85721, USA}
\affiliation{Department of Astronomy, University of Arizona, Tucson, 
  AZ 85721,USA}

%%%%%%%%%%%%%%%%%%%%%%%%%%%%%%%%%%%%%%%%%%%%%%%%
\begin{abstract}
  Hybrid hadron-quark equations of state that give rise a third family
  of stable compact stars have been shown to be compatible with the
  LIGO-Virgo event GW170817. Stable configurations in the third family
  are called hybrid hadron-quark stars. The equilibrium stable hybrid
  hadron-quark star branch is separated by the stable neutron star
  branch with a branch of unstable hybrid hadron-quark stars. The
  end-state of these unstable configurations has not been studied,
  yet, and it could have implications for the formation and existence
  of twin stars -- hybrid stars with the same mass as neutron stars
  but different radii. We modify existing hybrid hadron-quark
  equations of state with a first-order phase transition in order to
  guarantee a well-posed initial value problem of the equations of
  general relativistic hydrodynamics, and study the dynamics of
  non-rotating or rotating unstable twin stars via 3-dimensional
  simulations in full general relativity. We find that unstable twin
  stars naturally migrate toward the hadronic branch. Before settling
  into the hadronic regime, these stars undergo (quasi)radial
  oscillations on a dynamical timescale while the core bounces between
  the two phases. Our study suggests that it may be difficult to form
  stable twin stars if the phase transition is sustained over a large
  jump in energy density, and hence it may be more likely that
  astrophysical hybrid hadron-quark stars have masses above the twin
  star regime. We also study the minimum-mass instability for hybrid
  stars, and find that these configurations do not explode, unlike the
  minimum-mass instability for neutron stars. Additionally, our
  results suggest that oscillations between the two Quantum
  Chromodynamic phases could provide gravitational wave signals
  associated with such phase transitions in core-collapse supernovae
  and white dwarf-neutron star mergers.
\end{abstract}

\date{\today} \maketitle
%\tableofcontents

%%%%%%%%%%%%%%%%%%%%%%%%%%%%%%%%%%%
%SUITABLE SYNONYMS FOR "PRESSURE INCREASE": Pressure
% enlargement, enhancement, amplification, inflation
%%%%%%%%%%%%%%%%%%%%%%%%%%%%%%%%%%%
\section{Introduction}
%%%%%%%%%%%%%%%%%%%%%%%%%%%%%%%%%%%

It is a truly exciting time for nuclear (astro)physics as a number of
nuclear physics experiments and instruments observing the cosmos are
providing orthogonal information on the dense nuclear matter equation
of state (EOS) (see~\cite{Horowitz:2019piw, Raithel:2019uzi,
  Baiotti:2019sew, Radice:2020ddv, Chatziioannou:2020pqz} for recent
reviews). A crucial open question about the nuclear EOS is whether or
not quark deconfinement takes place in the high-density environments
of compact stars and what the nature of this phase transition is
(see~\cite{Baym:2017whm} for a recent review). In principle, the
densities inside stable neutron stars may reach the threshold for
quark deconfinement~\cite{PhysRevLett.34.1353, Heiselberg_2000,
  Fukushima:2013rx}. If the surface tension between the hadronic and
quark phases is strong enough to support a first-order phase
transition with a large jump over energy density, then a new branch of
stable compact stars emerges that is called the ``third
family''~\cite{Gerlach68, Kampfer1981, Schertler2000, Glendenning2000,
  Alford2017,Ayvazyan2013,Blaschke2013,Benic2015,Haensel2016,Bejger2017,
  Kaltenborn2017,Alvarez-Castillo2017,Abgaryan2018,Maslov2018,Blaschke2018,Alvarez-Castillo2019}. The
first family of stable compact objects are the white dwarves, and the
second family stable neutron stars. The third family of stars are
hybrid hadron-quark stars, i.e., they posses a quark core surrounded
by a hadronic shell. Twin stars are hybrid hadron-quark stars with the
same gravitational mass as neutron stars but more compact. Increased
attention has recently been paid to hybrid EOSs in the context of
constraining the neutron star EOS especially in the context of the
LIGO-Virgo event GW170817 (see, e.g.,\cite{Paschalidis:2017qmb,
  Bejger_2017,Montana:2018bkb, Bozzola:2019tit,
  Irving:2019yae,Bauswein:2020aag, Chatziioannou_Han2020,
  Shahrbaf2020, Abbott_2020GW170817,pereira2020tidal}
and~\cite{Baym:2017whm,Horowitz:2019piw} for reviews).

The monumental observation of a likely binary neutron star (BNS)
merger in both the electromagnetic (EM) and gravitational wave (GW)
spectrum (GW170817)~\cite{FirstDirectGW, GW170817_GRB170817,
  multimessGW170817} has led to a number of first constraints on the
hadronic nuclear EOSs from multimessenger observations (see,
e.g.,~\cite{Bauswein:2017vtn, Shibata:2017xdx,
  Margalit2017,Paschalidis:2017qmb, Abbott:2018exr, De_2018,
  Annala:2017llu, AnnalGordaKurkelVuor,Raithel:2018ncd,
  Rezzolla_2018Mmax, Ruiz:2017due, Radice:2018pdn,Capano:2019eae}).
Constraints on hadronic nuclear EOSs have also been placed from
observations of low mass x-ray binaries~\cite{Lattimer:2013hma,
  Ozel:2015fia, Ozel:2016oaf, Miller:2016pom}.  Additionally,
NICER~\cite{doi:10.1117/12.2231304} recently started to place
constraints on the dense matter EOS~\cite{Raaijmakers_2019,
  Miller_2019, Bogdanov_2019I, Bogdanov_2019II}.

Despite these efforts there still remain uncertainties in the dense
matter EOS above the nuclear saturation density. For instance, most
analyses of GW170817 were centered on EOSs which contain only hadronic
degrees of freedom but it is possible that at least one of the binary
components could have contained quark degrees of freedom as first
discussed in~\cite{Paschalidis:2017qmb} (see also \cite{Burgio2018,
  Drago2018, Hanauske2018, Montana2018,
  DePietri:2019khb,Annala:2019puf} and references therein). In fact,
some studies find that hybrid hadron-quark EOSs may be favored by
GW170817 over purely hadronic EOSs~\cite{Paschalidis:2017qmb,Han_2019,
  Essick:2019ldf}\footnote{Henceforth we refer to EOSs which include
  both hadron and quark degrees of freedom as ``hybrid EOSs'', and
  stars that contain both hadronic and quark phases as ``hybrid
  stars''. Hybrid stars with the same mass as neutron stars are
  referred to as ``twin stars.''}. Furthermore, several of the
existing constraints on the nuclear EOS depend on a number of
assumptions. Finally, an important caveat of some existing constraints
is that these either become less restrictive or do not apply when one
allows for hybrid EOSs~\cite{Paschalidis:2017qmb,Han_2019,
  Bozzola:2019tit}.  Of course whether hybrid stars (HSs) exist or not
will require additional observations and theoretical studies.

In light of the important information hybrid EOSs introduce when
constraining the nuclear EOS, it is crucial to better understand HSs,
including their dynamics. However, our understanding of hybrid EOSs in
the context of dynamical scenarios is currently in its infancy because
only a limited number of studies have been performed in general
relativity~\cite{Lin:2005zda, Abdikamalov:2008df, Dimmelmeier:2009vw, Bejger:2009zz,
  Haensel2016,Weih:2019xvw,Bauswein:2020aag,
  Chatziioannou_Han2020,Shahrbaf2020,Most:2018eaw, Hanauske:2018eej, Most:2019onn,
  Dexheimer:2019mhh,
  DePietri:2019khb,PhysRevLett.122.061102,bucciantini2019formation,Gieg:2019yzq,
  Weih_2020}. 

Equilibrium HSs in the third family are expected to be stable against
radial perturbations because they satisfy the turning point
theorem~\cite{1981ApJ...249..254S,1988ApJ...325..722F,Schiffrin:2013zta,Prabhu:2016pei}.
However, as in the case of equilibrium neutron stars and white
dwarves, where the stable neutron star and stable white dwarf branches
are separated by an unstable branch~\cite{Shapiro1983}, stable
third-family stars and neutron stars are separated by a branch of
unstable HSs. In particular, the unstable branch is in the twin star
regime. Thus, an important question concerns the fate of twin stars in
the unstable branch. If unstable twin stars naturally migrate to the
hadronic branch, then it may be difficult to form stable twin stars in
nature, because stable twin stars and neutron stars in current model
EOSs do not differ very much in their properties. If that is the case,
then a stable (proto-)neutron star that receives a strong
perturbation, e.g., following core collapse or a white dwarf--neutron
star merger, could temporarily migrate into the unstable twin-star
branch, but it would finally settle into the stable hadronic branch,
and therefore not form stable twin stars. Thus, collapsing stars or
even white dwarf--neutron star mergers might preferentially form
neutron stars. On the other hand, if unstable twin stars tend to
migrate toward the stable third-family branch, then it is possible
that stable twin stars can form in nature. What is more, the nature of
this instability could provide hints into the type of GW signatures
one could expect from events that can form twin stars, such as core
collapse supernovae (CCSN) ~\cite{Takahara:1988yd, Gentile:1993ma,
  Nakazato:2008su, Hempel_2009pe, Yasutake:2009kj}, the merger of a
white-dwarf with a neutron
star~\cite{Paschalidis:2009zz,Paschalidis:2010dh,Paschalidis:2011ez,Blaschke:2008na,
  Bejger:2011bq,Drago:2015dea} or the accretion induced collapse of a
white dwarf~\cite{Alvarez-Castillo:2019apz}. Finally, it is well known
that neutron stars have a minimum-mass instability, whose outcome is a
spectacular explosion~\cite{1977ApJ...215..311C, 1984SvAL...10..177B,
  Colpi1989ApJ...339..318C, Colpi:1990fe,1998A&A...334..159S}.  Thus,
it is interesting to explore what would happen to a twin star if by
some process, e.g., an eccentric black hole--twin star encounter, it
was brought near or slightly below the twin star minimum mass limit.

As a first step toward understanding which way the scales may tip --
neutron star or stable twin star or something else, we design hybrid
EOSs that give rise to a third family of compact objects, while taking
care that a well-posed initial value problem is guaranteed. We then
perform 3-dimensional, hydrodynamic simulations in full general
relativity to investigate the fate of unstable branch twin stars. We
consider different EOSs and a variety of non-rotating and rotating
twin star models along the unstable branch as well as different
initial perturbations. We find that unstable twin stars naturally
migrate toward stable neutron stars. The unstable configurations can
be momentarily driven away from the stable neutron star branch by
depleting the pressure. When driven away from the hadronic branch, the
stars undergo radial oscillations on a dynamical timescale while
bouncing between phases but ultimately settle into the hadronic
branch.  Rotating models also undergo these oscillations, allowing for
the possibility of GW signals. We find that the GWs associated with
oscillating rotating models may be detectable by future GW
observatories out to the Andromeda galaxy. When coupled with the
detection of GWs from potential progenitor systems of these types of
stars, such as CCSN~\cite{Takahara:1988yd, Gentile:1993ma,
  Nakazato:2008su, Hempel_2009pe, Yasutake:2009kj} or white
dwarf-neutron star (WDNS)
mergers~\cite{Paschalidis:2009zz,Paschalidis:2010dh,Paschalidis:2011ez,Blaschke:2008na,
  Bejger:2011bq,Drago:2015dea}, it is possible to expect a signal
characteristic of the evolution of unstable branch hybrid stars
corresponding to strong oscillations between the hadronic and quark
phases.  Explosions associated with minimum-mass
instability~\cite{1977ApJ...215..311C, 1984SvAL...10..177B,
  Colpi1989ApJ...339..318C, Colpi:1990fe,1998A&A...334..159S} were not
observed for minimum mass hybrid stars in our study.

The outline of the present work is as follows. In Sec.~\ref{sec:EOSID}
we summarize the EOSs we consider and detail our construction of
initial data. Sec.~\ref{sec:methods} includes a discussion of our
evolution methods and the diagnostics used to monitor the
simulations. In Sec.~\ref{sec:res} we discuss the ultimate fate of
unstable branch twin stars as we vary the initial perturbations and
EOSs. Additionally, in Sec.~\ref{sec:res} we study the minimum-mass
instability in the context of hybrid stars. In
Sec.~\ref{sec:discussion} we discuss the associated gravitational
radiation and the fate of rotating hybrid stars in the context of
constant rest mass equilibrium sequences.  As an additional
exploration of possible transitions between branches, we consider the
evolution of stable hybrid stars, which we present in
App.~\ref{app:stable_hybrids}.  We conclude in
Sec.~\ref{sec:conclusion} and point out future avenues of
investigation. Throughout this work we adopt geometrized units, where
$c=G=1$, unless otherwise noted.

\section{Equations of state}\label{subsec:EOS}
\begin{figure}
  \centering
\includegraphics{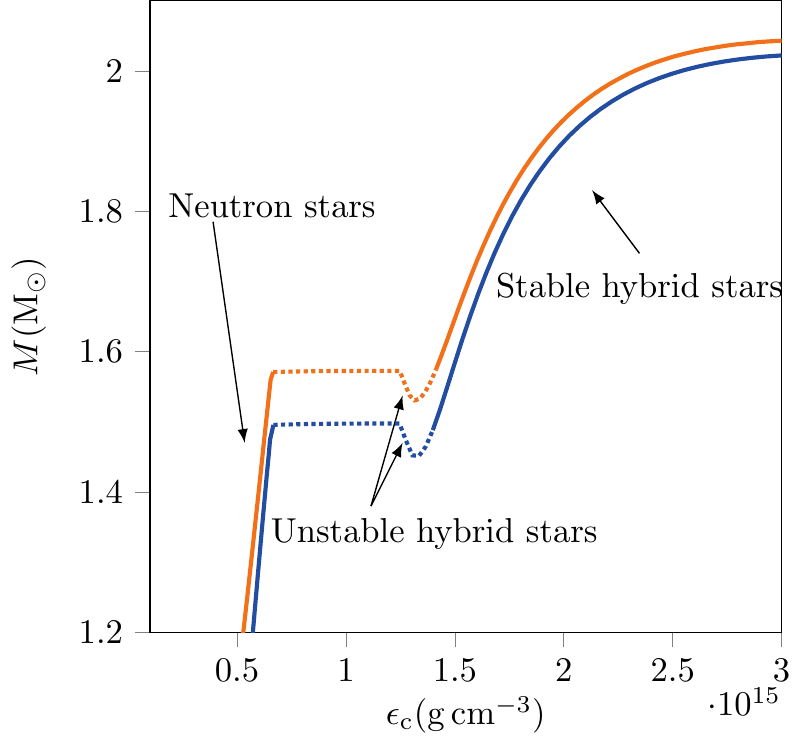}
\caption{ Constant angular momentum sequences depicting the
  gravitational mass $M$ as a function of the central energy density
  $\epsilon_c$ for EOS $\text{T9}$ of~\cite{Bozzola:2019tit} (labeled
  as EOS $\text{T9}_0$ in this work).  We show the non-rotating
  sequence (lower blue line) and a sequence where $cJ/GM_\odot^2=1.0$
  (upper orange line). Along each sequence we highlight the segments
  wherein twin stars roughly reside using dotted lines. We point out different
  segments of each sequence corresponding to stable neutron stars,
  stable hybrid stars, and unstable hybrid stars.}
  \label{fig:Jseq_T9_0}
\end{figure}

In this section we discuss the EOSs considered in this work. The EOSs
were chosen such that they are representative of the diversity of EOSs
treated in~\cite{Paschalidis:2017qmb,Bozzola:2019tit}. Current
constraints on the dense matter EOS allow for high-density quark
deconfinement phase transitions. Hybrid EOSs vary widely and lead to a
wide range of observable properties (see~\cite{Baym:2017whm} and
references therein for a review of viable models of hadron-quark
hybrid EOSs).  A full exploration of the space of astrophysically
consistent hybrid EOSs is beyond the scope of the present work, and
our focus is on hybrid EOSs with phase transitions with a large jump
in energy density, such that a third family of stable compact objects
emerges.

Before we proceed further, in Fig.~\ref{fig:Jseq_T9_0} we present a
gravitational mass--central energy density plot, to show the branches
of stable and unstable compact objects for the T9 EOS
in~\cite{Bozzola:2019tit}, which exhibits a first-order phase transition.
The plot shows regions of the sequences corresponding to stable
neutron stars, stable hybrid stars, and unstable hybrid stars for
non-rotating stars (lower blue line), and a constant-angular momentum
sequence with $cJ/GM_\odot^2=1.0$ (upper orange line).  We also
highlight segments of the sequences wherein twin stars roughly reside
using dotted lines.  According to the turning-point criterion for
stability~\cite{1988ApJ...325..722F, CST92, CST94a, CST94b}, along
sequence of stars of constant entropy $S$, and constant angular
momentum $J$, instability arises when
\begin{equation}\label{eq:sec_instab}
\dfrac{\partial M}{\partial \epsilon_{\rm c}} \Bigg \rvert_{J,S} \leq 0,
\end{equation}
where $M$ is the gravitational or Arnowitt-Deser-Misner (ADM) mass and
$\epsilon_{\rm c}$ is the central energy density.  The unstable twin
star branches are shown by the arrows in Fig.~\ref{fig:Jseq_T9_0},
where Eq.~\eqref{eq:sec_instab} is satisfied.

\subsection{Base EOSs with first-order phase transitions}

We focus on two representative EOSs from each of the two classes of
hybrid EOSs studied in~\cite{Paschalidis:2017qmb}
and~\cite{Bozzola:2019tit}.  Namely, we consider EOSs based on A4 and
T9, following the naming convention introduced
in~\cite{Bozzola:2019tit}; in~\cite{Paschalidis:2017qmb} the A4 and T9
models are labeled as ACB4 and ACS-II $j{=}0.9$, respectively. Since
we modify these original EOSs, in this work we label as $\text{A4}_0$
and $\text{T9}_0$ the original EOSs designated as A4 and T9
in~\cite{Bozzola:2019tit}, respectively.  Both EOSs correspond to
zero-temperature matter in beta equilibrium. Each EOS incorporates
hadronic (quark) degrees of freedom below (above) a threshold
transition energy density density, $\epsilon_{\rm tr}$. %% The
%%matching
%% in pressure between the two phases is performed via Maxwell
%% construction, corresponding to a first-order phase transition.
%% The onset of the phase transition occurs at an energy density of
%% $\epsilon_{\text{tr}}^{\text{A4}_0} \approx
%% \SI{5.6e14}{\g\per\cm\cubed}$ and $\epsilon_{\text{tr}}^{\text{T9}_0}
%% \approx \SI{6.58e14}{\g\per\cm\cubed}$ for EOSs $\text{A4}_0$ and
%% $\text{T9}_0$, respectively.\\

Among the different features between the EOSs are the parametrization
of the pressure in the different phases. EOS $\text{A4}_0$ includes a
parametrization of the pressure as a 4-segment piecewise polytrope,
\begin{equation}\label{eq:A4_EOS}
P = \kappa_i \rho_0^{\Gamma_i},
\end{equation}
where the index $i$ corresponds to the segment of the EOS and runs
from 1 to 4 in the case of $\text{A4}_0$, $\rho_0$ is the rest-mass
density, and $\kappa_i$ ($\Gamma_i$) is the polytropic constant
(adiabatic exponent) corresponding to segment $i$.  The specific values
for $\kappa_i$ and $\Gamma_i$, as well as the values of number
densities which demarcate each segment $n_{i}$, are listed in Tab.~II
of~\cite{Paschalidis:2017qmb}. EOS $\text{T9}_0$ includes a hadronic
phase EOS calculated using density functional
theory~\cite{Colucci:2013pya} and the quark phase EOS is calculated
using the MIT bag
model~\cite{Witten1984,Haensel1986,Alcock1986,Zdunik2000}. The
pressure in the quark phase for EOS $\text{T9}_0$ is parametrized
assuming a constant sound speed $c_{\rm s} $ as
\begin{equation}\label{eq:T9_EOS}
  P(\epsilon) =
  \begin{cases}
    P_{\text{tr}}                  & \epsilon_{\text{tr}}^{\text{T9}_0} \le \epsilon \le 1.9 
    \epsilon_{\text{tr}}^{\text{T9}_0}\,, \\
    P_{\text{tr}} + c_{\rm s}^2(\epsilon - 1.9 \epsilon_{\text{tr}}^{\text{T9}_0}) & \epsilon \ge 
    1.9\epsilon_{\text{tr}}^{\text{T9}_0}\,,
  \end{cases}
\end{equation}
where $c_{\rm s}=1.0$, $\epsilon$ is the energy density, $P_{\text{tr}}$ is the pressure of the hadronic 
phase at $\epsilon_{\text{tr}}^{\text{T9}_0}$, and the relativistic sound speed is defined 
as
\begin{equation}\label{eq:sound_speed}
c_{\rm s}^2 \equiv \dfrac{\partial P}{\partial \epsilon}.
\end{equation}
As shown in~\cite{Paschalidis:2017qmb,Bozzola:2019tit}, a key
difference between EOSs $\text{A4}_0$ and $\text{T9}_0$ is that the
former is an example of an EOS that gives high-mass ($\sim 2M_\odot$)
twin stars, while the latter results in low-mass ($\sim 1.5M_\odot$)
twin stars. Another key difference between EOSs $\text{A4}_0$ and
$\text{T9}_0$ is the response of their equilibrium configurations to
rotation. For EOS $\text{A4}_0$, sequences of increasing angular
momentum undergo a relatively large increase in mass for models with
central energy density $\epsilon_{\text{tr}}^{\text{A4}_0} \lesssim
\epsilon_{\rm c} \lesssim 1.78 \epsilon_{\text{tr}}^{\text{A4}_0}$,
which is roughly the region corresponding to the phase transition.
This relative increase results in the maximum mass stable hybrid star
having a smaller mass than the maximum mass hadronic star at large
values of the angular momentum. On the contrary, for the $\text{T9}_0$
EOS there is a comparable increase in mass for models at all values of
the energy density, which results in the maximum mass hybrid star
having a larger mass than the maximum mass hadronic star at all values
of the angular momentum. Considering these key differences between
EOSs A4 and T9, we aim to qualitatively cover a considerable part of
the space of hybrid hadron-quark EOSs.

\subsection{The challenge of evolving EOSs with first-order phase transitions}

The original $\text{A4}_0$ and $\text{T9}_0$ treat the phase
transition region using a Maxwell construction. As a result, the EOSs
have constant pressure during the phase transition which implies that
the speed of sound vanishes ($c_{\rm s}=0$) for the corresponding
values of the energy density. This is a problem for fluid dynamics and
numerical evolutions, because when the sound speed vanishes the
principal symbol of the equations of relativistic fluid dynamics does
not possess a complete set of eigenvectors (as a straightforward check
of the principal matrix in~\cite{Font1994A&A...282..304F} can
demonstrate). Hence the system of partial differential equations is
only weakly hyperbolic~\cite{GustafssonKreissOliger}.  For quasilinear
partial differential equations weak hyperbolicity generally implies an
ill-posed initial value problem~\cite{GustafssonKreissOliger}, which
precludes stable numerical integration. Thus, EOSs with zero speed of
sound can result in unstable numerical integration. To overcome this
problem, we raise the value of the sound speed over the phase
transition region. However, the modification is done such that the
third family does not disappear as we explain in the following
subsection.

\subsection{Modified EOSs}\label{subsec:modEOSs}
One way to modify EOSs with first-order phase transitions so that the
sound speed does not vanish is to change the type of construction
method for matching the hadronic and the quark phases. This can be
achieved by the well-known Glendenning
construction~\cite{GlendenningPhysRevD.46.1274}, which leads to a
smooth variation of the pressure over the phase transition
region. However, the Glendenning and Maxwell constructions are two
limits of the more general, and perhaps more physical, ``pasta phase''
construction~\cite{Maslov:2018ghi}. Other ``smoothing'' procedures
were introduced in~\cite{Abgaryan2018}, demonstrating that the third
family of compact stars does not depend on the existence of a
first-order phase transition. Moreover, the modifications that were
presented in the phase transition region are smooth enough to be
represented with a piecewise polytrope. Additionally, piecewise
polytropic parameterizations are able to capture the dynamics of the
phenomena we are interested in, including the minimum-mass neutron
star instability~\cite{1993ApJ...414..717C}. This motivates the
approach that we follow in the present work that we next turn to.

Given that piecewise polytropes simplify numerical hydrodynamic
simulations, we first employ an 11-branch piecewise polytrope to fit
both the $\text{A4}_0$ and $\text{T9}_0$ underlying EOSs, such that
the pressure as a function of rest-mass density is described by
Eq.~\eqref{eq:A4_EOS}.  The first 4 branches of our fits are reserved
for the crust, and for convenience we adopt the piecewise polytropic
representation of the SLy~\cite{Douchin_2001} EOS, provided in Tab.~II
of~\cite{Read2009}. For a given EOS, we match our piecewise polytropic
representation with the piecewise polytropic SLy parametrization at
the lowest possible energy density at which the two EOSs intersect. We
ensure that the pressure in our resulting EOSs is
monotonically increasing with rest-mass density. We have also verified
that the choice of crust EOS leaves the equilibrium stellar
configurations used in our set of initial data practically the same as
when using the baseline EOS tables.

\begin{table}[htb]\label{tab:EOS_props}
  \centering
  \caption{Properties of piecewise polytropic representations for the
    EOSs used in this work. We list the polytropic constants
    $\kappa_i$ (in cgs units determined by Eq.~\eqref{eq:A4_EOS}),
    adiabatic exponents $\Gamma_i$, and segment-dividing rest-mass
    densities $\rho_{0,i}$ (in units of $\SI{e15}{\g\per\cm\cubed}$)
    for each segment $i$.  We list the sound speed at the start of the
    phase transition $c_{\rm s}^{\rm tr}$ (in units of the speed of
    light $c$) and the rest-mass densities corresponding to the start
    and end of the phase transition transition $\rho_{0, \rm tr}^i$
    and $\rho_{0,\rm tr}^f$, respectively (in units of
    $\SI{e15}{\g\per\cm\cubed}$).  Note that the information for
    segments $1-4$ correspond to the SLy EOS~\cite{Read2009}.  The
    polytropic information for EOS $\text{A4}_0$ can be found in
    Tab.~II of~\cite{Paschalidis:2017qmb} in the entry corresponding
    to EOS ACB4.}
    \begin{tabular}{l | c c c c | c | c c }\hline  \hline
EOS & $i$ & $k_i$ & $\Gamma_i$ & $\rho_{0,i}$ & $c_{\rm s}^{\rm tr}$ 
%& $M_{\rm max}^{\rm had}$ & $M_{\rm max}^{\rm hyb}$ 
& $\rho_{0, \rm tr}^i$& $\rho_{0, \rm tr}^f$\\ \hline
$\text{A4}_0$  & --   & -- & -- & -- & 0.0 & 0.57 & 1.0 \\ \hline
A4& 1   & $6.11\times 10^{+12}$ & 1.584 & -- & 0.097 & 0.53 & 0.91 \\
  & 2   & $9.54\times 10^{+14}$ & 1.287 & $2.44\times 10^{-8}$ & & \\
  & 3   & $4.79\times 10^{+22}$ & 0.062 & $3.78\times 10^{-4}$ &  &\\
  & 4   & $3.59\times 10^{+13}$ & 1.359 & $2.63\times 10^{-3}$ &  &\\
  & 5   & $6.50\times 10^{+15}$ & 1.186 & 0.02 &  & \\
  & 6   & $6.62\times 10^{+03}$ & 2.060 & 0.05 &   &  \\
  & 7   & $5.40\times 10^{-04}$ & 2.559 & 0.16 &   &       \\
  & 8   & $3.59\times 10^{-38}$ & 4.921 & 0.29 &   &       \\
  & 9   & $1.15\times 10^{+32}$ & 0.200 & 0.53 &  &       \\
  & 10  & $1.65\times 10^{-25}$ & 4.000 & 0.91 &   &       \\
  & 11  & $2.09\times 10^{-10}$ & 3.000 & 1.27 &   &       \\ \hline
$\text{T9}_0$  & --  & -- & --    & --    & 0.0   & 0.66 & 1.0 \\ \hline
T9         & 1   & $6.11\times 10^{+12}$ & 1.584 & -- & 0.1 & 0.60 & 1.08 \\
           & 2   & $9.54\times 10^{+14}$ & 1.287 & $2.44\times 10^{-8}$ &  & & \\
           & 3   & $4.79\times 10^{+22}$ & 0.062 & $3.78\times 10^{-4}$ &  & & \\
           & 4   & $3.59\times 10^{+13}$ & 1.359 & $2.63\times 10^{-3}$ &  & & \\
           & 5   & $3.75\times 10^{+06}$ & 1.870 & 0.04 &       &       & \\
           & 6   & $1.43\times 10^{-03}$ & 2.536 & 0.14 &       &       &       \\
           & 7   & $1.02\times 10^{-15}$ & 3.377 & 0.29 &       &       &       \\
           & 8   & $3.03\times 10^{+30}$ & 0.3   & 0.60 &       &       &       \\
           & 9   & $2.52\times 10^{-49}$ & 5.56  & 1.08 &       &       &       \\
           & 10  & $1.16\times 10^{-18}$ & 3.535 & 1.38 &       &       &       \\
           & 11  & $9.91\times 10^{-02}$ & 2.427 & 1.88 &       &       &       \\ \hline
\end{tabular}
 \end{table}

%%%%%%%%%%%%%%%%%%%%%%%%%%%%%%%%%%%

%%%%%%%%%%%%%%%%%%%%%%%%%%%%%%%%%%%

Beyond the point at which we match with the SLy crust, we fit the
underlying EOSs using the remaining 7 branches. Our fitting algorithm
follows that of~\cite{Read2009}, which minimizes the relative error in
the pressure between the underlying EOSs and their corresponding
fits. There are two key sets of parameters which determine the optimal
fit of an arbitrary tabulated EOS using piecewise polytropes.  The
first set of parameters is the rest-mass densities which demarcate the
boundaries between neighboring polytropic segments $\rho_{0,i}$.  The
second set of parameters corresponds to the polytropic constants
$\kappa_{i}$ and adiabatic exponents $\Gamma_{i}$ which provide a fit to
the underlying EOS in question. In the following, we briefly discuss
the algorithm used to determine these two sets of parameters. We begin
by evenly dividing the range of rest-mass densities at which we fit
the EOS beyond the crust into log-equispaced intervals. This division
of rest-mass density serves as an initial guess for the optimal set of
$\rho_{0,i}$.  To avoid interpolation where possible, we ensure that
the set of dividing rest-mass densities $\rho_{0,i}$ corresponds to
points in the tabulated underlying EOS. The remainder of the algorithm
is carried out to optimize the set of $\rho_{0,i}$, where we focus on
the optimization of one polytropic segment at a time:
\begin{enumerate}
	\item Given the set of dividing rest-mass densities $\rho_{0,i}$, we use the 
	tabulated EOS to evaluate the corresponding set of dividing pressures
	$P_{i}$. 
	\item We determine the polytropic constants and adiabatic exponents for 
	the polytropic fit using the following expressions:
		\begin{enumerate}
		\item The adiabatic exponents are determined, assuming continuity 
		between segments, as
		\begin{equation}
		\Gamma_i = \dfrac{\log\left( P_i / P_{i-1}\right)}{\log\left( \rho_{0,i} / \rho_{0,i-1}\right)}
		\end{equation}
		\item The polytropic constants are then determined as
		\begin{equation}
		\kappa_i = \dfrac{P_i}{\rho_{0,i}^{\Gamma_i}}.
		\end{equation}
		\end{enumerate}
	We highlight that the set of polytropic constants $\kappa_i$ are not independent of the 
	adiabatic exponents $\Gamma_i$. Once we set the transition densities $\rho_{0,i}$, we
	then use these and the underlying EOS to determine the adiabatic exponents 
	(step 2a. above), which in turn provide the polytropic constants (step 2b. above).
	\item We evaluate the root-mean-square (RMS) error between a
          linear interpolation of the tabulated pressure and the
          polytropic fit to the pressure at values of the rest-mass
          density $\rho_0$ which span the entire range of rest-mass
          densities in the table. In particular, the RMS error is calculated as
	\begin{equation}\label{eq:RMS}
	\text{RMS}[P] \equiv \sqrt{ \sum_n^N \dfrac{ \left(P_{\text{tab}}(\rho_{0,n}) - P_{\text{poly}}
	(\rho_{0,n}) \right)^2}{N}},
	\end{equation}
	where $\rho_{0,n}$ corresponds to the elements of a list of
        $N$ rest-mass densities. We typically choose $N=1000$ and
        choose the rest-mass densities $\rho_{0,n}$ such that they are
        log-equispaced between the minimum and maximum values of the
        rest-mass density in the table.  $P_{\text{tab}}$ is the
        pressure corresponding to the tabulated underlying EOS
        linearly interpolated to the rest-mass density $\rho_{0,n}$,
        and $P_{\text{poly}}$ is the pressure corresponding to the
        piecewise polytropic fit at $\rho_{0,n}$.  We employ linear
        interpolation because the underlying table is dense and to
        avoid oscillations arising from discontinuous pressure
        derivatives around sharp features of the EOS such as the start
        and end of the phase transition.  Eq.~\eqref{eq:RMS} allows us
        to determine the RMS error over the entire EOS for a
        particular choice of $\rho_{0,i}$ corresponding to the current
        EOS segment.
	\item We then vary $\rho_{0,i}$ for the current EOS segment
          until the RMS error is minimized. Once this is done, we
          focus on the neighboring segment and repeat the algorithm
          starting at step 1 above, until all segments have been
          chosen such that the RMS error is minimized for each segment
          separately. Once we have updated the values of all
          $\rho_{0,i}$ that constitutes one iteration in the fitting
          algorithm.
\end{enumerate}

\begin{figure}
  \centering
\includegraphics{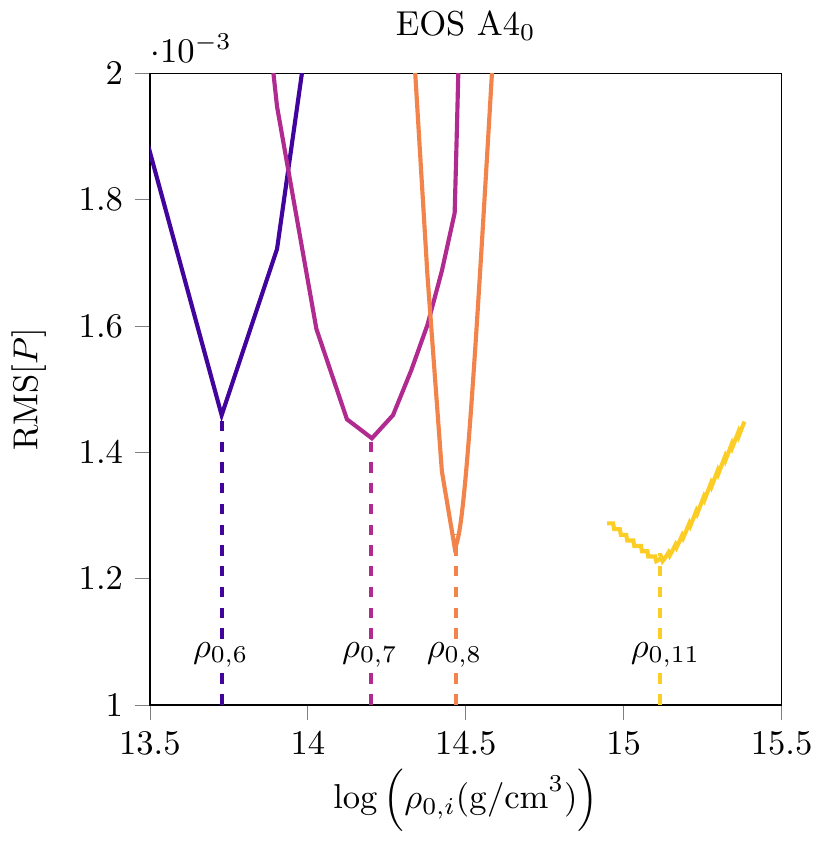}
\caption{Example of the root-mean-square error in the pressure as a function of the choice 
for the  dividing rest-mass density corresponding to segments 
6, 7, 8, and 11 (dark blue, magenta, orange, and yellow lines, 
respectively) of the polytropic fit for EOS $\text{A4}_0$. For each segment, we mark 
the value of $\rho_{0,i}$ that results in the minimum RMS error (and are thus optimal 
choices) with a dashed vertical line of the same color.}
  \label{fig:RMS}
\end{figure}

We perform multiple iterations of the fitting algorithm until the
minimum RMS error saturates and no longer decreases, thereby ensuring
that the set of $\rho_{0,i}$ is optimal. In Fig.~\ref{fig:RMS} we
show, as an example, the RMS error corresponding to the first
iteration of the fitting algorithm in the case of the $\text{A4}_0$
EOS.  For each polytropic segment, depicted by different color lines
in Fig.~\ref{fig:RMS}, the RMS error clearly reaches a minimum at the
best choice of $\rho_{0,i}$ for that segment.  We note that, as
suggested by the curves in Fig.~\ref{fig:RMS}, we only optimize the
dividing rest-mass densities corresponding to segments 6, 7, 8, and 11
because segments 1-4 correspond to the SLy crust, segment 5
corresponds to the point at which the crust matches the high density
EOS, and segments 9 and 10 correspond to the start and end of the
phase transition, respectively. As such, $\rho_{0,6}, \rho_{0,7},
\rho_{0,8}$, and $\rho_{0,11}$ are the only members of $\rho_{0,i}$
which should be optimized, and all others are left fixed.

In fitting each EOS with piecewise polytropes, we have full control of
the adiabatic index (and thus the sound speed) during the phase
transition $\Gamma_{\text{tr}}$. Note that fixing neighboring members
of $\rho_{0,i}$ at the points corresponding to the start and end of
the phase transition ensures that we fit that region with a single
polytrope. We experimented with several values of
$\Gamma_{\text{tr}}$, and chose values slightly above the lowest one
for which the evolution of stable equilibrium hybrids stars did not
exhibit numerical instabilities. We discuss the evolution of such
models in App.~\ref{app:stability_resolution}. The value chosen for
$\Gamma_{\text{tr}}$ corresponds to $c_{\rm s}^{\text{tr}} = 0.097$
($c_{\rm s}^{\text{tr}}=0.1$) at rest-mass density $\rho_0 =
\SI{5.3e14}{\g\per\cm\cubed}$ ($\rho_0 =
\SI{6.0e14}{\g\per\cm\cubed}$) for EOS A4 (T9).  In
Tab.~\ref{tab:EOS_props} we list the polytropic constants $\kappa_i$,
adiabatic exponents $\Gamma_i$, and dividing rest-mass densities
$\rho_{0,i}$ that provide the fit to our nonzero sound speed versions
of the $\text{A4}_0$ and $\text{T9}_0$ EOSs.  Henceforth, we use the
labels A4 and T9 to correspond to the piecewise polytropic, nonzero
sound speed EOSs used in this work and summarized in
Tab.~\ref{tab:EOS_props}. Note that in Tab.~\ref{tab:EOS_props}, the
polytropic information is listed such that for $\rho_0 > \rho_{0,i}$,
the pressure is given by $P=\kappa_i \rho_0^{\Gamma_i}$.  Hence, the
first entry for $\rho_{0,i}$ is left blank because the first segment
of the EOS may begin at any rest-mass density below $\rho_{0,2}$.

\section{Initial Data \label{sec:EOSID}}

\begin{figure*}
  \centering
\includegraphics{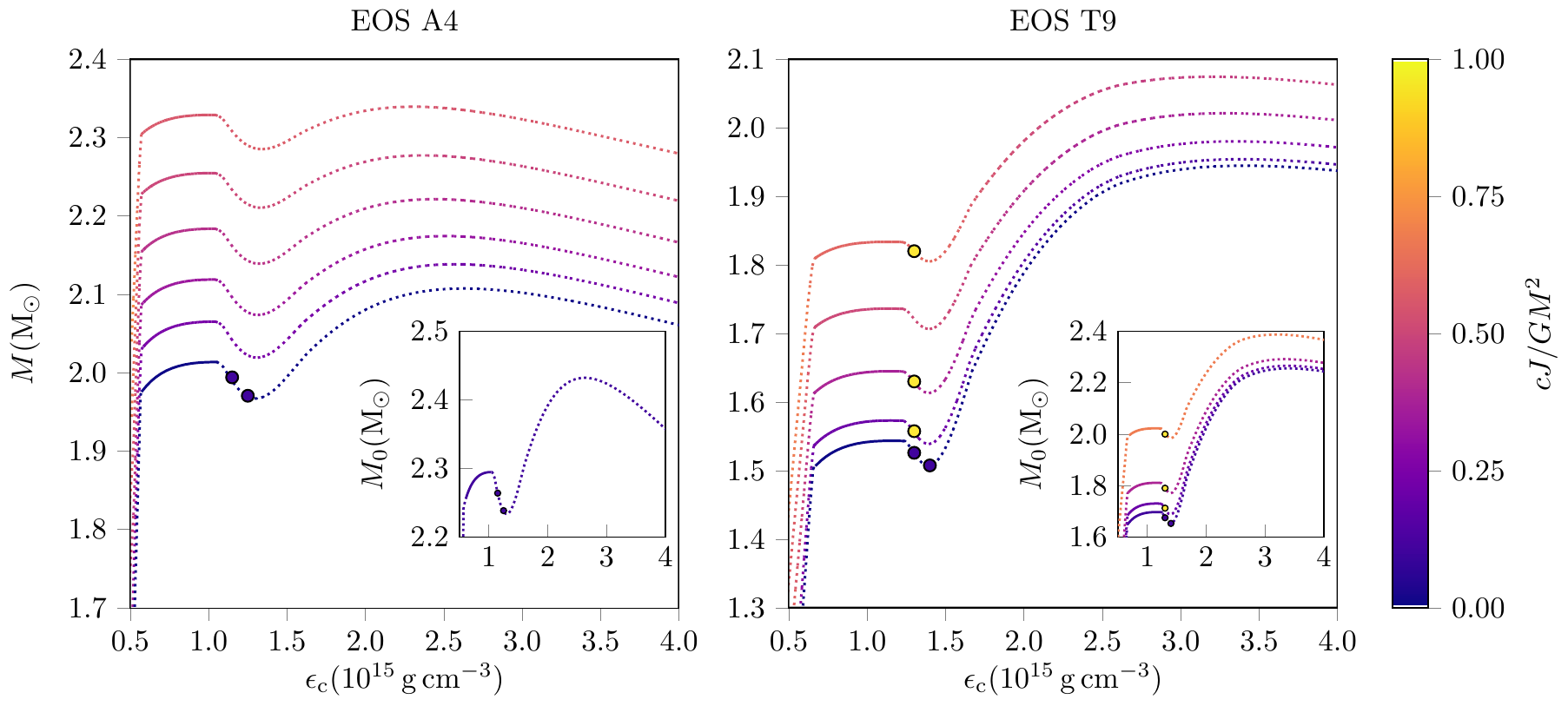}
\caption{\textit{Left panel:} Sequences of constant angular momentum,
  and zero entropy depicting the gravitational mass $M$ as a function
  of the central energy density $\epsilon_c$ for EOS A4. The colorbar
  corresponds to the dimensionless spin $a = cJ/GM^2$ along each
  sequence.  For each sequence we highlight the additional stable
  models introduced by modifying the polytropic fit to the underlying
  $\text{A4}_0$ EOS using solid line segments.  We mark the models
  that comprise our set of initial data using circles (dark blue for
  non-rotating and bright yellow for rotating). In the inset we show
  the rest mass $M_0$ as a function of central energy density
  $\epsilon_{\rm c}$ for sequences which correspond to our initial
  data for this EOS. \textit{Right panel:} Same as left panel but for
  EOS $\text{T9}$.  }
  \label{fig:Jseq_A4}
\end{figure*}

In this section we discuss the methods for generating our initial data
and the properties of the initial configurations we considered.

\subsection{Methods}

We construct initial data with the code of~\cite{CST92, CST94a,
  CST94b}, which solves Einstein's equations coupled to the equations
of hydrostationary equilibrium for a perfect fluid assuming
stationarity and axisymmetry. The code adopts the following spacetime
metric~\cite{CST94a} (see~\cite{Paschalidis:2016vmz} for a review of
other line elements used in the literature)
\begin{equation}\label{eq:IDmetric}
ds^2 = -e^{\gamma+\rho}dt^2+e^{2\alpha}(dr^2+r^2d\theta^2)+e^{\gamma-\rho}r^2\sin^2\theta(d
\phi-\omega 
dt)^2,
\end{equation}
where $r$, $t$, $\theta$, and $\phi$ correspond to the usual spherical coordinates, and 
$\gamma$, $\rho$, $\omega$, and $\alpha$ are metric potentials which are functions of 
$r$ and $\theta$ only. The perfect fluid stress-energy tensor is given by
\begin{equation}
T^{ab} = \rho_0 h u^a u^b + P g^{ab},
\end{equation}
where $u^a$ is the fluid four velocity and $h$ is the specific enthalpy,
given by
\begin{equation}
h = 1 + \varepsilon + \dfrac{P}{\rho_0},
\end{equation}
where $\varepsilon$ is the specific internal energy . To close the system of equations, the EOSs 
described in Sec.~\ref{subsec:EOS} are supplied.

\subsection{Equilibrium configurations}

Using our EOSs A4 and T9 we build equilibrium configurations of
compact stars.  In Fig.~\ref{fig:Jseq_A4}, we show constant angular
momentum sequences for the A4 EOS (left panel) and T9 EOS (right
panel) for rigidly rotating stars.  As demonstrated in the figure, our
modified EOSs give rise to a third family of compact stars, including
branches of stable and unstable twin stars. In addition, the inclusion
of rotation results in a relative increase in mass for different
regions of the EOS fits, as previously described for the underlying
EOSs. More specifically, we find that the A4 EOS is most alike the
$\text{A4}_0$ EOS beyond the phase transition region, and results in a
close match of the maximum mass hybrid star with $\sim 0.2\%$ relative
difference in the mass, while it results in a $\sim 2\%$ difference in
the maximum mass star on the hadronic branch. On the other hand, the
T9 EOS most closely matches the $\text{T9}_0$ EOS below the
phase-transition region, resulting in a close match of the maximum
mass hadronic star with less than $0.1 \%$ relative difference in the
mass, while it results in a $\sim 4\% $ relative difference in the
maximum mass of the third-family branch. Thus, our fits provide good
approximations to the underlying EOSs without the shortcoming of a
zero sound speed over the phase transition region.

The key difference introduced into the EOSs by increasing the sound
speed is a change from a sharp, first-order phase transition to a
smoother transition between phases. As mentioned earlier, both sharp
and smooth phase transitions are consistent with a high-density quark
deconfinement phase transition, and each scenario has been considered
in the study of hybrid stars in the past~\cite{Hempel_2009pe,
  Yasutake:2009kj, Bhattacharyya_2010mg, Yasutake_2011mg}.  In the
context of the turning-point instability, increasing the sound speed
over the phase transition region leads to a nonzero first derivative
$M'(\epsilon_c)$ over the span of the transition, where before it was
0, and hence the $\text{A4}_0$ and $\text{T9}_0$ EOSs contained only
turning-point marginally unstable models. Compared to the original
$\rm A4_0$ and $\rm T9_0$, our A4 and T9 EOSs have an extended second
family of compact objects that in addition to neutron stars now also
includes stars whose maximum density is above the phase transition
threshold, and hence are hybrid stars. Moreover, the maximum mass
configuration in the second family is also a hybrid star. This is a
result of the mixed hadron-quark phase introduced by modifying of the
original EOSs.

%$\epsilon_{\text{tr}}^{\text{A4}_0} \approx \SI{5.72e14}{\g\per\cm\cubed}$ and 
%$\epsilon_{\text{tr}}^{\text{T9}_0} \approx \SI{6.56e14}{\g\per\cm\cubed}$ 
\begin{table}[htb]\label{tab:model_props}
  \centering
  \caption{Properties of the initial unstable equilibrium twin star
    configurations considered in this work. For each model we show the
    central energy density $\epsilon_c$ and rest-mass density
    $\rho_{0, \rm c}$ in units of $\SI{e15}{\g\per\cm\cubed}$, the ADM
    mass $M$ and rest mass $M_0$ in units of $\SI{}{M_\odot}$, the
    compactness $C\equiv GM/R_cc^2$ (where $R_c$ is the equatorial
    circumferential radius), the ratio of polar to equatorial radii
    $r_p / r_e$, and dimensionless spin $a \equiv cJ / G M^2$. For
    each model, in the last two columns we also list the central
    rest-mass density (in units of $\SI{e15}{\g\per\cm\cubed}$) of the
    lower (second family) and higher density (third family) stable
    stars with rest mass equal to the unstable configuration. The last
    line in the table corresponds to the minimum mass unstable hybrid
    twin star with the T9 EOS, which does not have a corresponding
    higher-density configuration on the stable third-family branch.  }
  \begin{tabular}{l c c c c c c c | c c}\hline  \hline
EOS & $\epsilon_c$ & $\rho_{0, \rm c}$ & $M$ & $M_0$ & $C$ & $\frac{r_p}{r_e}$ & $a$ 
& $\rho_{0,c, \rm low}$ & $\rho_{0,c \rm high}$ \\\hline
A4  & 1.15 & 0.97 & 1.99 & 2.27 & 0.21 & 1.00 & 0.00 & 0.61 & 1.21 \\
    & 1.25 & 1.04 & 1.97 & 2.24 & 0.22 & 1.00 & 0.00 & 0.53 & 1.13 \\ \hline
T9  & 1.30 & 1.12 & 1.53 & 1.68 & 0.17 & 1.00 & 0.00 & 0.73 & 1.25 \\ \hline
T9  & 1.30 & 1.12 & 1.56 & 1.71 & 0.17 & 0.96 & 0.20 & 0.70 & 1.25 \\
    & 1.30 & 1.12 & 1.63 & 1.80 & 0.17 & 0.87 & 0.40 & 0.76 & 1.28 \\
    & 1.30 & 1.12 & 1.82 & 2.00 & 0.17 & 0.68 & 0.60 & 0.68 & 1.27 \\ \hline
T9  & 1.40 & 1.19 & 1.50 & 1.65 & 0.17 & 1.00 & 0.00 & 0.62 & -- \\ \hline
\end{tabular}
 \end{table}

In Fig.~\ref{fig:Jseq_A4} we mark with filled circles the points along
the constant angular momentum sequences which correspond to our set of
initial data. This set samples a range of masses of twin stars on the
unstable branch, with the lowest and highest mass stars having
$M\approx 1.5 M_\odot$, and $M\approx 2.0 M_\odot$, respectively. Some
of our models have rest masses which are greater than that of the
corresponding maximum mass hadronic stars $M_{\rm max}^{\rm had}$,
leading to the possibility that they settle into the stable hybrid
configurations in the second family introduced after smoothing the
phase transition transition (see insets in each panel of
Fig.~\ref{fig:Jseq_A4} for constant $J$ sequences which show the rest
mass $M_0$ as a function of the energy density).  In
Fig.~\ref{fig:Jseq_A4} we highlight these additional stable models
using solid line segments along each constant angular momentum
sequence. We discuss the possibility of settling into these additional
stable configurations in Sec.~\ref{sec:res}.  Our rotating models
(marked by yellow filled circles in the right panel of
Fig.~\ref{fig:Jseq_A4}) are chosen by considering constant angular
momentum sequences which contain stable stars on both the hybrid and
hadronic branches and locating models which satisfy
Eq.~\eqref{eq:sec_instab}.  We include relevant properties for the set
of initial data in Tab.~\ref{tab:model_props}.

\subsection{Initial perturbations}
\label{sec:init_pert}
\begin{table}[htb]\label{tab:sim_summary}
  \centering
  \caption{Main set of cases considered in this work. For each model
    we list the model name (see text body for model naming
    convention), the EOS, the central energy density in units of
    $\SI{e15}{\g\per\cm\cubed}$, the dimensionless spin $a=cJ / G
    M^2$, and the value of the perturbation parameter $\xi$ in
    Eq.~\eqref{eq:p_pert}. The final entry, model $\rm
    MMT9_{1.4}^{-2\%}$, corresponds to the evolution of the
    minimum-mass non-rotating hybrid star for the T9 EOS in the case
    where $2\%$ of the rest mass is removed at the start of the
    simulation.}
  \begin{tabular}{l | c c c c c }\hline  \hline
Model & EOS & $\epsilon_c$ & $a$ & $\xi$  \\ \hline
$\rm A4_{1.15}^{-0.01}$        & A4   & 1.15   & 0.0  & -0.01  \\
$\rm A4_{1.15}^{0}$            & A4   & 1.15   & 0.0  &  0.0   \\
$\rm A4_{1.15}^{+0.01}$        & A4   & 1.15   & 0.0  &  0.01  \\ \hline
$\rm A4_{1.25}^{-0.01}$        & A4   & 1.25   & 0.0  & -0.01  \\
$\rm A4_{1.25}^{0}$            & A4   & 1.25   & 0.0  &  0.0   \\
$\rm A4_{1.25}^{+0.01}$        & A4   & 1.25   & 0.0  &  0.01  \\ \hline
$\rm T9_{1.3}^{-0.01}$         & T9   & 1.3    & 0.0  & -0.01  \\
$\rm T9_{1.3}^{0}$             & T9   & 1.3    & 0.0  &  0.0   \\
$\rm T9_{1.3}^{+0.01}$         & T9   & 1.3    & 0.0  &  0.01  \\ \hline
$\rm T9_{1.3}^{0.0}a_{0.2}$    & T9   & 1.3    & 0.2 &  0.0   \\
$\rm T9_{1.3}^{-0.01}a_{0.2}$  & T9   & 1.3    & 0.2 & -0.01  \\ \hline
$\rm T9_{1.3}^{0.0}a_{0.4}$    & T9   & 1.3    & 0.4 &  0.0   \\
$\rm T9_{1.3}^{-0.01}a_{0.4}$  & T9   & 1.3    & 0.4 & -0.01  \\ \hline
$\rm T9_{1.3}^{0.0}a_{0.6}$    & T9   & 1.3    & 0.6 &  0.0   \\
$\rm T9_{1.3}^{-0.01}a_{0.6}$  & T9   & 1.3    & 0.6 & -0.01  \\ \hline
$\rm MMT9_{1.4}^{-2\%}$        & T9   & 1.4    & 0.0  &  0.0 \\ \hline
\end{tabular}
 \end{table}
 
For each configuration presented in Tab.~\ref{tab:model_props} we consider the effects of  
seeding a small perturbation at the start of the simulation. We focus on quasi-radial 
perturbations which 
are seeded by perturbing the pressure at $t=0$ everywhere in the star. The form of the 
pressure perturbation is
\begin{equation}\label{eq:p_pert}
P(t=0, \mathbf{x}) \longrightarrow (1+\xi) P(t=0, \mathbf{x}), 
\end{equation}
where $\mathbf{x}$ indicates the spatial coordinates, and $\xi$ can be
either positive or negative in cases where we add or deplete pressure,
respectively.  Our main set of simulations is summarized in
Tab.~\ref{tab:sim_summary}, and consists of 16 cases. The naming
convention for the models presented in Tab.~\ref{tab:sim_summary} is
as follows. The model name begins with the EOS label, followed by a
superscript corresponding to the value of the perturbation parameter
in Eq.~\eqref{eq:p_pert} and a subscript corresponding to the value of
the initial central energy density (in units of
$\SI{e15}{\g\per\cm\cubed}$) which determines the model's location on
the unstable branch.  For rotating configurations, the model name is
followed by the letter `a' (corresponding to the dimensionless spin)
\begin{equation}\label{eq:spin}
a \equiv \dfrac{cJ}{GM^2},
\end{equation} 
and a subscript corresponding to the value of $a$. For example, the
non-rotating model corresponding to the A4 EOS with $\epsilon_{\rm c}
= \SI{1.15e15}{\g\per\cm\cubed}$ under a $1\%$ positive pressure
perturbation is labeled $\rm A4^{+0.01}_{1.15}$, while the rotating
model corresponding to the T9 EOS with $\epsilon_{\rm c} =
\SI{1.3e15}{\g\per\cm\cubed}$ under a $1\%$ negative pressure
perturbation and spin $a=0.4$ is labeled $\rm
T9^{-0.01}_{1.3}a_{0.4}$.  In Tab.~\ref{tab:sim_summary} we also list
the simulation $\rm MMT9_{1.4}^{-2\%}$, which corresponds to the
evolution of the minimum mass non-rotating hybrid star for the T9 EOS,
where $2\%$ of the initial rest mass is stripped at the start of the
simulation (here we use $2\%$ and non 0.02 in the case label to
indicate that it is not a pressure perturbation).  Along with the
simulations presented in Tab.~\ref{tab:sim_summary} we also consider
two simulations at varying grid resolutions corresponding to model
$\rm A4^{-0.01}_{1.25}$, a set of simulations corresponding to stable
hybrid stars used to asses the validity of each EOS, two simulations
corresponding to different size initial perturbations for model $\rm
T9_{1.3}^{-0.01}$, three simulations corresponding to stable branch
hybrid twin stars used to consider the possibility of migration from
the stable branch to the unstable regime, and several simulations
corresponding to the lower and higher density equilibria with the same
rest mass as that of particular models in Tab.~\ref{tab:sim_summary}
as comparison points. Our stable twin star and resolution studies are
presented in App.~\ref{app:stability_resolution}. Our study of
dynamical migration from the stable hybrid branch is presented in
App.~\ref{app:stable_hybrids}.

\section{Evolution methods}
\label{sec:methods}
In this section we describe the basic methods used in evolving the
initial data outlined in Sec.~\ref{sec:EOSID}. We describe the
evolution code, the grid hierarchy, and detail the different
diagnostics used.
  
\subsection{Evolution code}\label{subsec:evol}
To evolve the hydrodynamics and spacetime we use the code
of~\cite{Duez:2005sg, ILGRMHD_2010, Etienne:2015cea}, which operates
within the {\tt Cactus} framework~\cite{Cactus} and employs {\tt
  Carpet}~\cite{Schnetter:2003rb, Schnetter:2006pg} for mesh
refinement. The code solves the Einstein equations using the
Baumgarte-Shapiro-Shibata-Nakamura formulation~\cite{Shibata:1995we,
  Baumgarte:1998te} within the 3+1 formalism. Our gauge choice
consists of ``1+log" slicing for the lapse~\cite{Bona:1994dr} and the
``Gamma-freezing" condition for the shift in first-order
form~\cite{Alcubierre:2001vm, Alcubierre:2002kk, Etienne:2007jg}.  The
temporal evolution uses a fourth-order Runge-Kutta scheme with a
Courant-Friedrichs-Lewy factor of 0.5.  The fluid variables are
evolved in flux-conservative form adopting high-resolution
shock-capturing methods~\cite{Etienne:2010ui,Etienne:2011re}. Our code
is compatible with piecewise polytropic representations of realistic,
cold, beta-equilibrated EOSs. We validate our approach to modify
hybrid hadron-quark equations of state by evolving stable branch
hybrid star models. We present the results of these evolutions in
App.~\ref{app:stability_resolution}. The code has been thoroughly
tested in the past and demonstrated to be convergent. Of relevance to
this work are the convergence tests in~\cite{Paschalidis:2010dh}.

\subsection{Grid hierarchy}  
\label{subsec:grid}
For all evolutions in this work, we construct evolution grids, using
fixed mesh refinement~\cite{Schnetter:2003rb, Schnetter:2006pg},
consisting of 7 nested boxes. The half-side length of the finest level
is set to $1.5 R_{c}$ (where $R_{c}$ is the initial hybrid star
equatorial circumferential radius), and all subsequent levels have
half-side length equal to twice that of the adjacent finer level. The
canonical resolution used in our study is set such that the finest
level contains at least 64 grid-points per $R_{c}$, so that the finest
canonical grid spacing is given by $\Delta x_1 = R_{c}/64$. All other
levels have grid spacings $\Delta x_n = 2 \Delta x_{n-1}$, where $n\in
(2,7)$ is the level number with larger $n$ meaning coarser level. In
addition, we consider higher resolution simulations to assess
convergence and invariance of our results with resolution. For higher
resolution runs, we employ grids which are $1.25$ and $1.5$ times
finer than the canonical-resolution grid, which we label the medium-
and high-resolution cases, respectively. Note that there are at least
80 and 96 grid-points per $R_{c}$ for the medium- and high-resolution
simulations, respectively. To reduce computational cost we employ
reflection symmetry across the equatorial plane.  To avoid
singularities when converting the initial stellar solutions from
spherical polar to the Cartesian coordinates used in the evolution, we
shift the grid points in the y-direction by $\Delta y = 0.001$ (in
units where $(G/c^2)\SI{e15}{\g\per\cm\cubed}$ equals 1), so that the
origin is avoided. Such coordinate shifts have been shown to have a
negligible effect on the dynamics of relativistic
stars~\cite{Espino:2019ebx}. 

\subsection{Diagnostics}
We monitor several diagnostics to assess different aspects of the
evolution including the evolution of the rest-mass density, the $L_2$
norm of the Hamiltonian and momentum
constraints~\cite{Etienne:2007jg}, and global conservation laws such as
total rest-mass, total ADM mass, and angular momentum
conservation. We also track the boundary between the hadronic and
quark phases, which we define as the locus of points where the
rest-mass density $\rho_{0}$ equals the value corresponding to the
onset of the phase transition for a given EOS.

Although our initial configurations are in equilibrium, rotating
models undergoing quasi-radial oscillations generate GWs.  To investigate this, we extract GWs using the
Newman-Penrose formalism~\cite{Newman:1961qr, Penrose:1962ij}, in
particular focusing on $s=-2$ spin-weighted spherical harmonic
decompositions of the Newman-Penrose scalar $\Psi_4$. The coefficients
of the spin-weighted decomposition are labeled $\Psi_4^{l,m}$, where
$l$ and $m$ are the usual degree and order for the spherical
harmonics. In all cases we focus on the dominant quasi-radial $l=2$,
$m=0$ mode. We extract $\Psi_4^{l,m}$ from the numerical solution at
fixed concentric spheres with increasing coordinate radii
$r_{\rm ex} = \eta M$, where $\eta \in
\left\{40,50,60,80,90,100\right\}$. For a suitable comparison to the
GWs of similar systems~\cite{Dimmelmeier:2009vw, Haensel2016}, we
compute the gravitational wave strain $h$ from
\begin{equation}
\Psi_4 = \ddot h_+ - i \ddot h_\times.
\end{equation}
adopting the fixed-frequency integration (FFI)~\cite{Reisswig_2011GWs}. 
The visualizations and GW analysis presented in this work were carried out 
using the \texttt{kuibit} Python package~\cite{kuibit}.
%% FFI allows the calculation of the
%% strain as
%% \begin{equation}\label{eq:FFI}
%% h(t) = \text{FT}^{-1}\left[\dfrac{\text{FT}\left[\ddot h\right]}{f^2} \right],
%% \end{equation}
%% where $\text{FT}$ and $\text{FT}^{-1}$ denote the Fourier transform and its inverse, 
%% respectively. 
%% We employ a low-frequency cutoff in $f$ of $f_0=\SI{140}{kHz}$%% , such that the 
%% denominator in Eq.~\eqref{eq:FFI} is set to $f_0$ for $f<f_0$. The low-frequency cutoff is 
%% chosen by considering a family of strains $h(t;f_0)$ and setting $f_0$ such that the 
%% relative difference in strain between the maximum (high-frequency peak) and final 
%% (low-frequency) peak in $h$ roughly matches that in $\Psi_4$, thereby ensuring that the 
%% strain retains the same general shape as the underlying $\Psi_4$ signal. 

\section{Results}
\label{sec:res}
In this section we detail the results of the simulations listed in
Tab.~\ref{tab:sim_summary}. We first discuss our non-rotating models,
categorizing by the perturbation seeded at the beginning of the
simulations. We highlight the key features in the evolution and point
out the differences between the results for each EOS. We then present
the results for rotating models and highlight the key differences in
the dynamics introduced by rotation. Finally, we summarize the results
of our study of the hybrid star minimum-mass instability.

\subsection{Non-rotating models} \label{subsec:nonrot_res}
Our set of non-rotating models consists of the first 9 entries in
Tab.~\ref{tab:sim_summary}. The initial data for this set consists of
three models: two correspond to the A4 EOS, and one to the T9 EOS. The
initial data set also covers a range of central energy densities and
masses in the unstable twin star branch. In general, we find that all
non-rotating models in our study migrate toward the stable hadronic
configuration, i.e., the neutron star with the same rest mass but
lower maximum rest-mass density than the original
configuration. Generally, adding (removing) pressure initially, such
that $\xi>0$ ($\xi<0$), results in the configuration settling near the
hadronic branch on a shorter (longer) timescale than cases wherein no
perturbation was applied.  Depending on the EOS and central energy
density of the initial configuration, slight qualitative differences
arise between the evolutions.  However, \emph{all} non-rotating models
tend toward the hadronic branch (the second family of stable compact
objects), showing that the outcome of the instability is independent
of the perturbations we studied. In the following we consider the
effect of each perturbation separately.

\subsubsection{Equilibrium evolution} \label{subsubsec:equi_nonrot}
\begin{figure*}
  \centering
  \includegraphics{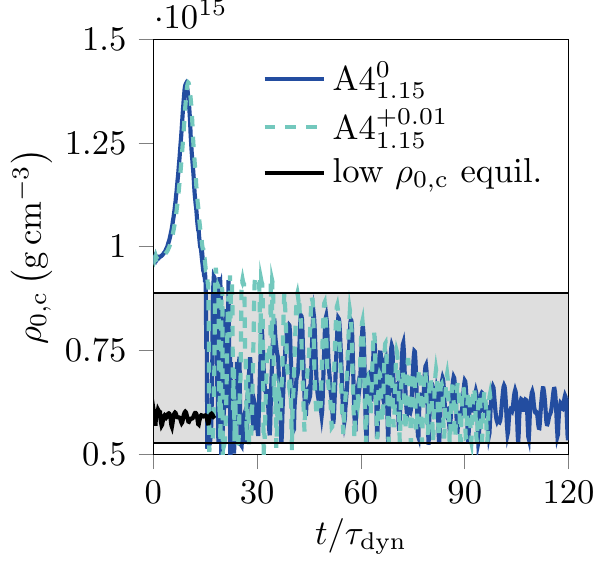}
  \includegraphics{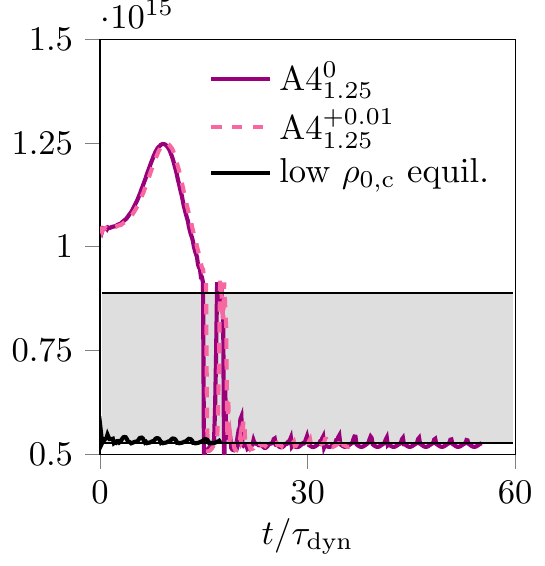}
  \includegraphics{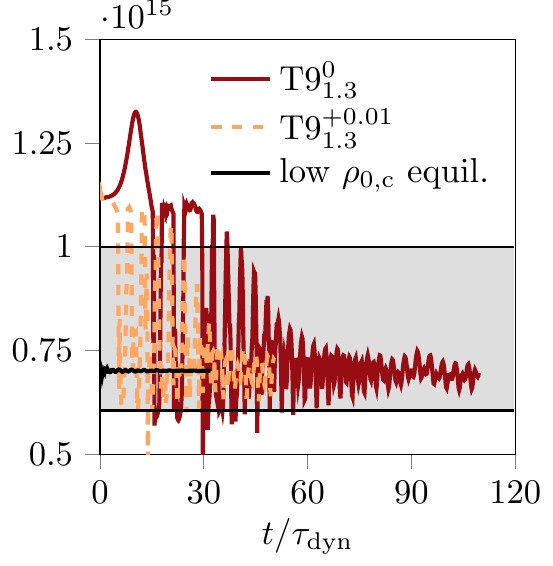}
  \caption{ \textit{Left panel:} Central rest-mass density $\rho_{0,
      \rm c}$ as a function of time (scaled by the dynamical time
    $\tau_{\rm dyn}$) for model $\rm A4_{1.15}^{0}$ under equilibrium
    evolution (solid lines) and model $\rm A4_{1.15}^{+0.01}$ with a
    positive pressure perturbation (dashed lines).  Also shown is
    $\rho_{0, \rm c}$ for the lower-density equilibrium model with the
    same rest mass as that of the initial configuration (solid black
    line). We use horizontal black lines to mark the start (lower
    lines) and end (upper lines) of the phase transition and a gray
    band to mark the densities corresponding to the phase transition
    region. \textit{Center panel:} Same as the left panel but for
    models $\rm A4_{1.25}^{0}$ and $\rm
    A4_{1.25}^{+0.01}$. \textit{Right panel:} Same as the left and
    center panels but for models $\rm T9_{1.3}^{0}$ and $\rm
    T9_{1.3}^{+0.01}$.}
  \label{fig:equi_inc_nonrot}
\end{figure*}

In Fig.~\ref{fig:equi_inc_nonrot} we present the central rest-mass
density as a function of time for the case of equilibrium evolution
for all non-rotating models in our study. Note that in
Fig.~\ref{fig:equi_inc_nonrot} the time is scaled by the dynamical
time, given by $\tau_{\rm dyn} = 1/\sqrt{\rho_{0,\rm max}(t=0)}$.  The
central rest-mass densities for models $\rm A4_{1.15}^{0}$, $\rm
A4_{1.25}^{0}$, and $\rm T9_{1.3}^{0}$ are depicted in the left,
center, and right panels of Fig.~\ref{fig:equi_inc_nonrot},
respectively, using solid lines. We mark the densities corresponding
to the phase transition region of the EOS using gray shaded regions in
Fig.~\ref{fig:equi_inc_nonrot}, such that central rest-mass densities
below the gray shaded regions correspond to pure hadronic
stars.

In the case of equilibrium evolution all non-rotating models in our
study undergo an initial increase in the rest-mass density
corresponding to contraction, which takes place early in the
evolution. At this point the size of the quark core saturates, the EOS
stiffens, and the models subsequently begin expanding. As the models
expand the value of the central rest-mass density eventually becomes
smaller than the threshold corresponding to the end of the phase
transition $\rho_{0, \rm tr}^f$ and quickly drops to a value $\rho_{0,
  \rm c} \lesssim \rho_{0, \rm tr}^i$ (see Tab.~\ref{tab:EOS_props}
for the values of $\rho_{0, \rm tr}^i$ and $\rho_{0, \rm tr}^f$ for
each EOS).  The expansion reverts into a momentary collapse as the
models begin to contract until $\rho_{0,\rm c} > \rho_{0,\rm tr}^i$,
at which point the central rest-mass density jumps to a point above
the end of the phase transition $\rho_{0, \rm c} \gtrsim \rho_{0, \rm
  tr}^f$. The models undergo several such bounces corresponding to
radial oscillations while the core oscillates between the hadronic and
quark phases. Eventually, the radial oscillations decay as the models
settle into approximately steady states at lower central rest-mass
densities.

The size and number of oscillations observed early in the evolution
likely depends on the position of each model along the unstable branch
(i.e., on the model's central energy density and rest mass). Of the
models considered, $\rm A4_{1.25}^{0}$ is the only one with rest mass
lower than the maximum rest mass purely hadronic star -- the
configuration with central rest-mass density just below the phase
transition threshold.  In other words, models $\rm A4_{1.15}^{0}$ and
$\rm T9_{1.3}^{0}$ have counterpart lower-density, stable equilibrium
hybrid star models with the same rest mass, which are in the second
family of compact objects (see solid segments of
Fig.~\ref{fig:Jseq_A4}). We find significant oscillations throughout
the evolutions of models $\rm A4_{1.15}^{0}$ and $\rm T9_{1.3}^{0}$
which peak inside the gray bands corresponding to the phase transition
region in Fig.~\ref{fig:equi_inc_nonrot}. However, we find a lack of
such long-term oscillations in model $\rm A4_{1.25}^{0}$, possibly
because this unstable twin star has a lower-density stable counterpart
with max density below that of the phase transition threshold. Such a
disparity in dynamics between unstable branch stars (whether or not
their central regions oscillate within the phase transition region)
may be tied to our EOSs where the quark deconfinement phase transition
is not a first-order transition. For sharp, first-order phase
transitions, all EOS models we are aware of produce unstable branch
twin stars with corresponding lower-density neutron stars which have
central rest-mass density lower than $\rho_{0, \rm tr}^i$.  If the
quark deconfinement phase transition is of first-order, the migration
of unstable branch hybrid stars toward the hadronic branch may closely
follow the evolution of model $\rm A4_{1.25}^{0}$. However, since the
pasta phase reconstruction may be more natural, there may be a
diversity in how unstable twin stars migrate toward the hadronic
branch. The $\rm A4_{1.15}^{0}$ and $\rm T9_{1.3}^{0}$ models both
undergo small but non-negligible late-time oscillations in the
rest-mass density of approximately $\sim 5\%$ in amplitude. We evolved
these models for more dynamical times than the $\rm A4_{1.25}^{0}$
model, which reveals that the amplitude of the rest-mass density
oscillations continues to decrease, albeit at a slower rate compared
to the initial oscillations.

For suitable comparisons we also consider the evolution of the stable
lower-density equilibrium model with the same rest mass as that of the
initial configuration of each model.  We depict the central rest-mass
density for these models with solid black lines in
Fig.~\ref{fig:equi_inc_nonrot}. In all cases the evolution of the
lower-density counterparts show small ($\lesssim 1\%$) oscillations in
the rest-mass density.  It is worth noting that the final central
rest-mass density of the lower-density counterparts agrees to within
$\sim 0.08\%$, $\sim 1.5\%$, and $\sim 0.1\%$ with the final central
rest-mass density of models $\rm A4_{1.15}^{0}$, $\rm A4_{1.25}^{0}$
and $\rm T9_{1.3}^{0}$, respectively. By the end of the evolutions it
is clear that all unstable models are very close to their
lower-density stable counterpart in the second family.

\subsubsection{Positive initial pressure perturbation} \label{subsubsec:inc_nonrot}
The evolution of $\rho_{0, \rm c}$ for the case of positive pressure
perturbations is shown by the dashed lines in
Fig.~\ref{fig:equi_inc_nonrot}.  We generally find that adding a
positive initial pressure perturbation results in evolutions that are
similar to those without initial perturbation. Depending on the EOS,
the key features of evolution may arise on a slightly shorter
timescale. For instance, where model $\rm T9_{1.3}^{0.0}$ settled into
a lower-density configuration with $\sim 5\%$ oscillations in the
rest-mass density after $t\approx 60\tau_{\rm dyn}$, we find that
adding pressure drives the initial configuration to a similar state by
$t\approx 30\tau_{\rm dyn}$. The timescale on which models $\rm
A4_{1.15}^{+0.01}$ and $\rm A4_{1.25}^{+0.01}$ settle to lower
densities is almost identical to models $\rm A4_{1.15}^{0}$ and $\rm
A4_{1.25}^{0}$, respectively.

Depending on the EOS, the early evolution in cases with positive
pressure perturbations show small differences to the cases of
equilibrium evolution. For models $\rm A4_{1.15}^{+0.01}$ and $\rm
A4_{1.25}^{+0.01}$ we observe an initial small expansion, which is
followed by contraction and subsequent evolution similar to the cases
wherein no perturbations were excited. However, model $\rm
T9_{1.3}^{+0.01}$ undergoes an initial expansion, without contraction,
which coincides with an initial decrease in the central rest-mass
density. The model continues expanding until the quark phase
disappears from the stellar center. The model then undergoes a number
of bounces and the evolution proceeds in a qualitatively similar
fashion to that of model $\rm T9_{1.3}^{0}$.

The differences in the early stages of evolution between model $\rm
T9_{1.3}^{+0.01}$ and those corresponding to the A4 EOS may be
attributed to the EOS stiffness in the quark phase immediately above
the phase transition region.  The first polytropic segment after the
phase transition region (listed in Tab.~\ref{tab:EOS_props} as segment
10 and 9 for EOSs A4 and T9, respectively) corresponds to the part of
the EOS that is sampled in the central region of these stars. As such,
the central regions of the initial configurations are significantly
stiffer for models built using the T9 EOS than for those built with
the A4 EOS.

\subsubsection{Pressure depletion} \label{subsubsec:dep_nonrot}
\begin{figure*}
  \centering
  \includegraphics{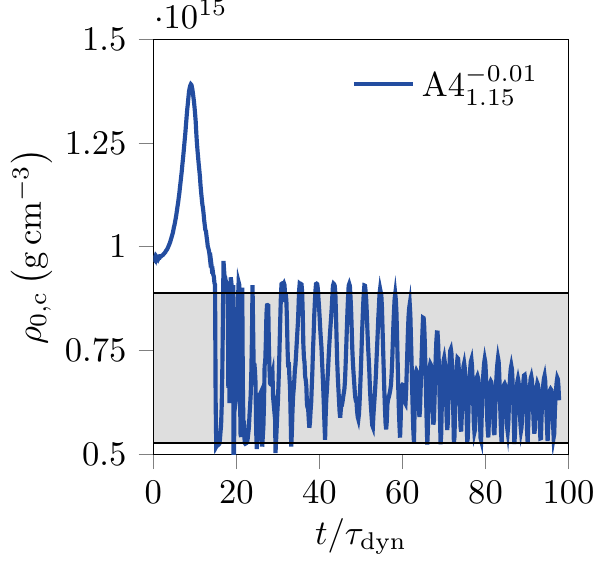}
  \includegraphics{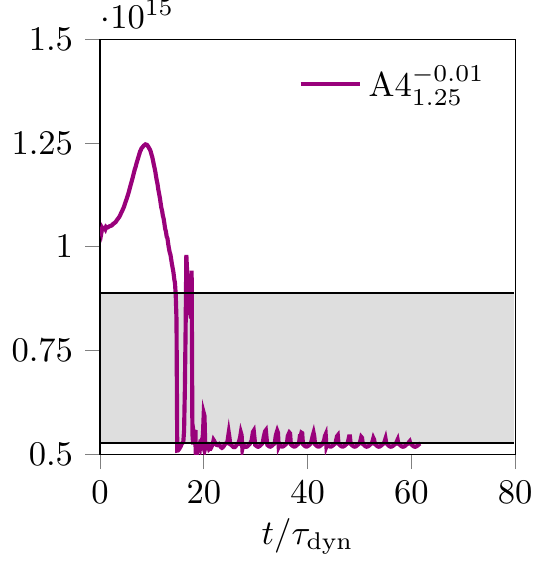}
  \includegraphics{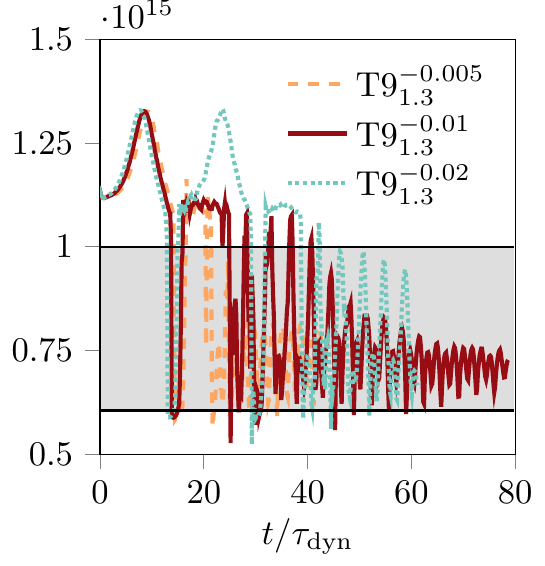}
  \caption{ \textit{Left panel:} Central rest-mass density $\rho_{0,
      \rm c}$ as a function of time (scaled by the dynamical time
    $\tau_{\rm dyn}$) for model $\rm A4_{1.15}^{-0.01}$ under a
    negative pressure perturbation.  We use horizontal black lines to
    mark the start (lower lines) and end (upper lines) of the phase
    transition and a gray band to mark the densities corresponding to
    the phase transition region.  \textit{Center panel:} Same as the
    left panel but for model $\rm A4_{1.25}^{-0.01}$.  \textit{Right
      panel:} Same as the left and center panels but for model $\rm
    T9_{1.3}^{-0.01}$. Also shown are $\rho_{0,\rm c}$ for model $\rm
    T9_{1.3}^{-0.005}$ with $0.5\%$ pressure depletion (light orange
    dashed line) and $\rm T9_{1.3}^{-0.02}$ with $2\%$ pressure
    depletion (blue dotted line).}
  \label{fig:dep_nonrot}
\end{figure*}
\begin{figure*}
  \centering
  \includegraphics{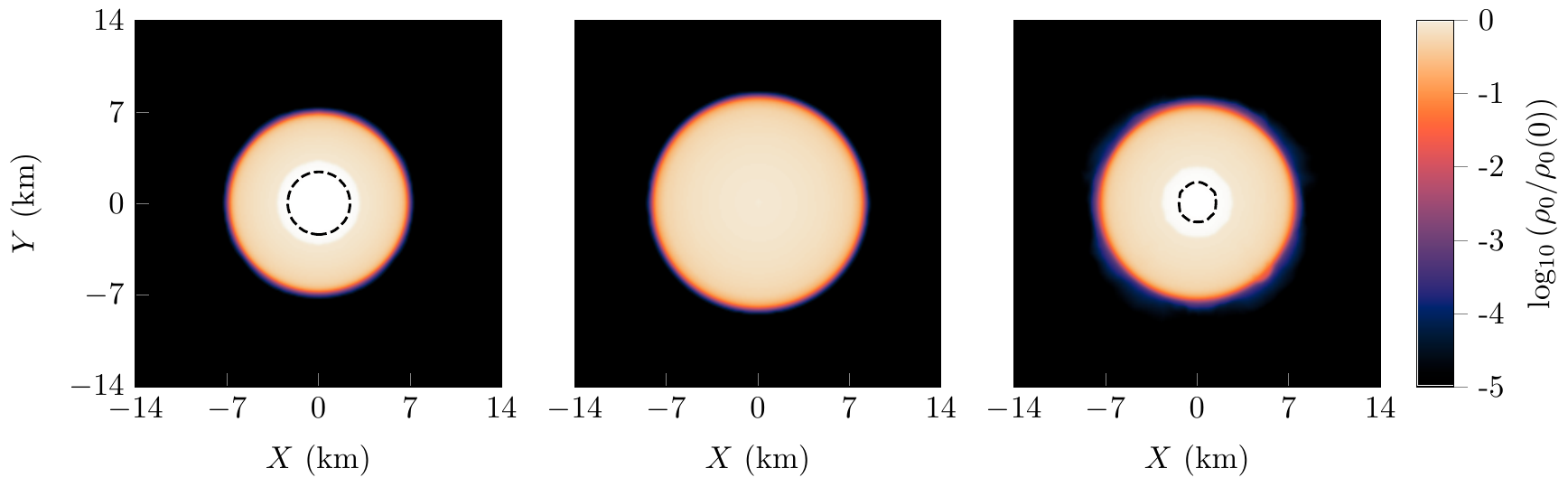}
  \caption{\textit{Left panel:} Equatorial snapshot of the rest-mass
    density for model $\rm A4_{1.15}^{-0.01}$ at time $t \approx
    10\tau_{\rm dyn}$, which corresponds to the peak of the central
    rest-mass density. We outline the region with rest-mass density
    larger than the end of the phase transition (corresponding to
    the quark core, shown in white) with a black dashed line.  
    At this stage the model exhibits a large quark core.
    \textit{Center panel:} Same as the left panel, but for time $t
    \approx 15\tau_{\rm dyn}$, which corresponds to the disappearance
    of the quark phase and the expansion of the model following the
    initial contraction.  \textit{Right panel:} Same as the left and
    center panels, but for time $t \approx 40\tau_{\rm dyn}$, which
    corresponds to a relative peak in the rest-mass density and the
    re-appearance of the quark core.}
  \label{fig:T9dep_snaps}
\end{figure*}

In this section we describe our results for an initial perturbation
that depletes ($\xi <0$) a small fraction of the pressure
everywhere. This is designed to test if the unstable configuration can
be pushed to the stable twin star counterpart. In
Fig.~\ref{fig:dep_nonrot} we show the central rest-mass density in the
case of pressure depletion. We generally find that relative to the
equilibrium evolution, pressure depletion tends to delay the timescale
on which some key features arise. Similarly to the cases of
equilibrium evolution and positive pressure perturbations, the models
considered under pressure depletion settle at lower central rest-mass
densities toward the second family branch. For all models under
pressure depletion we observe an initial increase in rest-mass
density. This initial increase in rest-mass density leads to the
growth of the quark core, such that a larger part of the core is above
the densities of the phase transition region, where the EOS stiffens
again. The stiffening of the EOS at these high densities leads to a
bounce, and the configurations enter an expansion phase, which results
in a decrease of the central rest-mass density. Once the central
rest-mass density falls below the critical value corresponding to the
end of the phase transition (upper solid black line in the left panel
of Fig.~\ref{fig:dep_nonrot}), it quickly drops to the rest-mass
density corresponding to the start of the phase of transition (lower
solid black line) and the quark core disappears.  As the core enters a
softer part of the EOS, the pressure support becomes weaker, and the
star quickly reverts into a momentary collapse. The cycle repeats and
the maximum rest-mass density exhibits strong oscillations as the
central region bounces between the quark and hadronic phases.  Similar
to cases wherein no explicit perturbation was excited, such
oscillations eventually decay and the models settle to lower central
rest-mass densities.  Depending on the properties of the initial
configurations (such as central energy density and mass), pressure
depletion may induce a larger number of such oscillations compared to
cases with no explicit perturbation.

The $\rm A4_{1.15}^{-0.01}$ model (left panel of
Fig.~\ref{fig:dep_nonrot}) undergoes strong bounces between the quark
and hadronic phases for the first $t\approx 60 \tau_{\rm dyn}$ of
evolution, while the oscillations in the central rest-mass density are
damped as the model settles to the lower-density stable equilibrium
model with the same rest mass as that of the initial configuration
(note that for models $\rm A4_{1.15}^{0.0}$ and $\rm
A4_{1.15}^{+0.01}$ the strong oscillations were damped after $t\approx
20 \tau_{\rm dyn}$). Model $\rm T9_{1.3}^{-0.01}$ (right panel of
Fig.~\ref{fig:dep_nonrot}) similarly exhibits strong oscillations for
longer ($t\approx 40 \tau_{\rm dyn}$) than models $\rm T9_{1.3}^{0.0}$
($t\approx 30 \tau_{\rm dyn}$) and $\rm T9_{1.3}^{+0.01}$ ($t\approx
20 \tau_{\rm dyn}$).  In the case of the $\rm A4_{1.25}^{-0.01}$ model
(center panel of Fig.~\ref{fig:dep_nonrot}) we find that the
configuration settles at the lower-density equilibrium model with the
same rest mass as that of the initial configuration on a comparable
(but slightly longer) timescale than the corresponding equilibrium
evolution case. Model $\rm A4_{1.25}^{-0.01}$ shows oscillations of a
comparable number, duration, and amplitude to model $\rm
A4_{1.25}^{0.0}$, suggesting that this initial configuration has the
highest propensity to migrate toward the stable second-family branch as
discussed in Sec.~\ref{subsubsec:equi_nonrot}. 

In Fig.~\ref{fig:T9dep_snaps} we show snapshots of equatorial contours
of the rest-mass density at key points corresponding to the evolution
of model $\rm A4_{1.15}^{-0.01}$. We outline the quark phase (shown in
white), with $\rho_{0} > \rho_{0, \rm tr}^f$, using black dashed
lines. The left panel of Fig.~\ref{fig:T9dep_snaps} depicts the
saturation of the quark core size during the initial contraction stage
of the model at $t\approx10\tau_{\rm dyn}$. The center panel of
Fig.~\ref{fig:T9dep_snaps} corresponds to the disappearance of the
quark core during the expansion stage which follows the initial
contraction stage at $t\approx15\tau_{\rm dyn}$. The right panel of
Fig.~\ref{fig:T9dep_snaps} depicts the time corresponding to a
local-in-time peak in the rest-mass density at $t\approx 40\tau_{\rm
  dyn}$.  All local-in-time maxima depicted in the left panel of
Fig.~\ref{fig:dep_nonrot} between $t\approx 20\tau_{\rm dyn}$ and
$t\approx 60\tau_{\rm dyn}$ are consistent with the right panel of
Fig.~\ref{fig:T9dep_snaps}. Fig.~\ref{fig:T9dep_snaps} shows that the
central region exhibits a quasi-periodic revival of the quark core as
the model undergoes radial oscillations. 
The evolutions of unstable branch hybrid stars with 
a negative initial pressure perturbation suggests the
ability of phase transitions giving rise to a third family to
drive strong oscillation cycles in these stars. This could be a
smoking gun signature of EOSs with sharp hadron/quark phase
transitions leading to a third family of compact objects (as has also
been suggested in~\cite{Hanauske:2018eej,Gieg:2019yzq}, for example,
following the merger of heavy neutron stars that form a remnant whose
density is above the threshold for the quark deconfinement phase
transition).

A common feature in all of the evolutions discussed thus far is that
the amplitude of density oscillations decreases as the models tend
toward a steady state, which suggests some form of energy
dissipation. For a qualitative understanding of the role that heating
may play in dissipating the stellar oscillations, we considered the
ratio of total to cold polytropic pressure $P_{\rm tot} \slash P_{\rm
  cold}$ to reveal the relative size of the thermal contribution
$P_{\rm th}$, computed through
  \begin{equation}\label{eq:p_tot}
  P_{\rm tot}  = P_{\rm cold} + P_{\rm th},
  \end{equation}
with a density cutoff of $\rho_0 = \SI{1e-3}{\rho_{\rm 0, max}}$.  In
all cases, we find that the models develop warm atmospheres (with over
$10\%$ thermal support), but that the bulk of the star remains cold
(with less than $1\%$ thermal pressure support). Our resolution study
showed that the oscillations are damped on approximately the same
timescale independent of resolution. However, we cannot reliably
conclude that the chief mechanism behind the damping of stellar
oscillations is shock heating and that some numerical dissipation is
not at play. We note that damping of such radial oscillations of stars
have been reported in other cases where migration from an unstable to
a stable branch can take place for neutron stars,
e.g.,~\cite{Guercilena:2016fdl}.

\subsubsection{Effect of initial perturbation amplitude}

To investigate the role of the amplitude of perturbations in
Eq.~\eqref{eq:p_pert}, we also consider perturbations which are half
as large ($\xi=-0.005$) and twice as large ($\xi=-0.02$) as that
considered for model $\rm T9_{1.3}^{-0.01}$, corresponding to $0.5\%$
and $2\%$ pressure depletion, respectively.  We ensure that for the
additional levels of pressure depletion the Hamiltonian and momentum
constraints remain small, and similar to the standard cases of
pressure depletion (see App.~\ref{app:stability_resolution} for a
discussion of the constraints with and without pressure
perturbations).  We present the central rest-mass density for these
cases with dashed orange ($\rm T9_{1.3}^{-0.005}$) and dotted blue
($\rm T9_{1.3}^{-0.02}$) lines in the right panel of
Fig.~\ref{fig:dep_nonrot}. We find that changing the size of the
perturbation affects the timescale on which the initial bounces
occur. Specifically, with larger pressure depletion ($\rm
T9_{1.3}^{-0.02}$), we find that the central rest-mass density reaches
its first maximum on a slightly shorter timescale compared to the
canonical pressure depletion case ($\rm T9_{1.3}^{-0.01}$). On the
other hand, lower pressure depletion ($\rm T9_{1.3}^{-0.005}$) results
in the first maximum being reached on a slightly longer timescale.
The size of the density perturbations does not strongly affect the
amplitude of the initial density maximum, but it affects the size of
late-time oscillations. We find that the amplitude of the density
oscillations for model $\rm T9_{1.3}^{-0.005}$ does not reach the
initial maximum after the first bounce (similar to model $\rm
T9_{1.3}^{-0.01}$), resulting in fewer and weaker bounces compared to
model $\rm T9_{1.3}^{-0.02}$. On the other hand, model $\rm
T9_{1.3}^{-0.02}$ reaches the peak value of $\rho_{0,\rm c}$ twice and
exhibits stronger oscillations deeper into the the evolution than
model $\rm T9_{1.3}^{-0.01}$. Ultimately, the oscillations for all
pressure depletion cases are damped as the models tend toward the
lower-density equilibrium models with the same rest mass as that of
the initial configurations. We find that the initial increase in
density is never large enough for the final configurations to reach
densities near the counterpart stable twin star. This is despite the
fact that the maximum density reached during the evolution exceeds the
central density of the counterpart configuration in the third
family. This is indicative of the natural propensity these unstable
solutions have to migrate toward the second-family branch.

\subsection{Rotating models}\label{subsec:rot_res}
  \begin{figure*}
  \centering
   \includegraphics{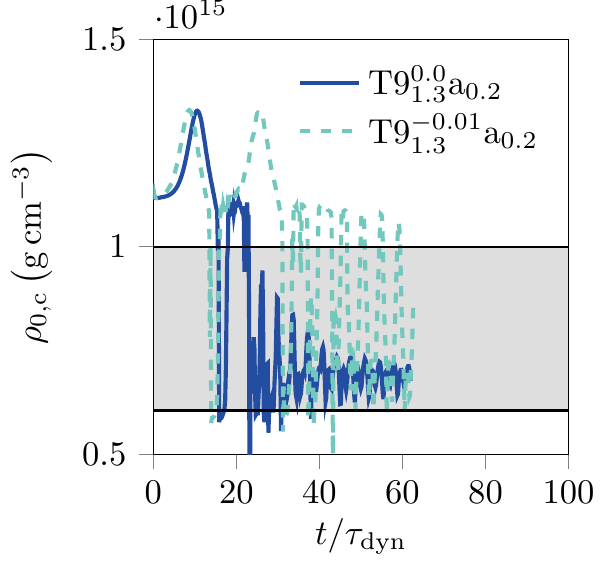}
   \includegraphics{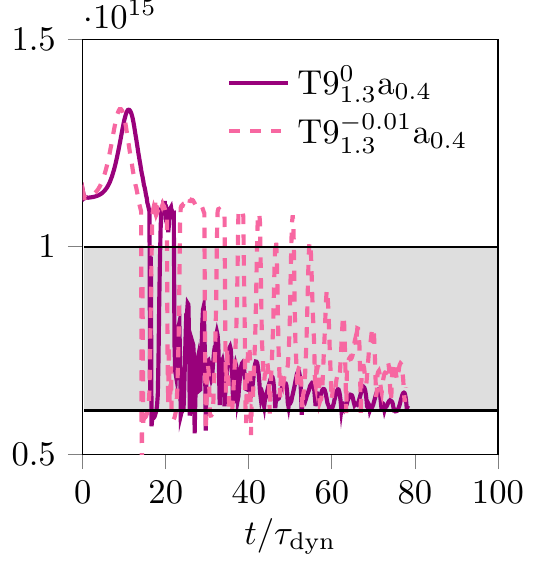}
   \includegraphics{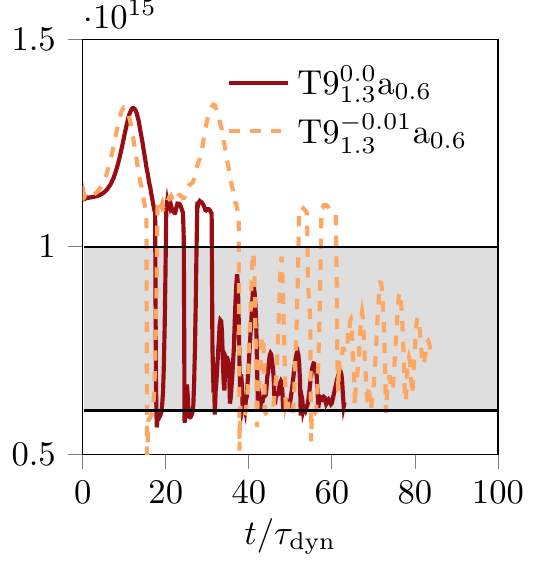}
  \caption{\textit{Left panel:} Central rest-mass density $\rho_{0,c}$
    as a function of time (scaled by the dynamical time $\tau_{\rm
      dyn}$) for models $\rm T9_{1.3}^{0}a_{0.2}$ with no initial
    perturbation (solid line) and $\rm T9_{1.3}^{-0.01}a_{0.2}$ with a
    negative initial pressure perturbation (dashed line). The lower
    and upper black horizontal lines correspond to the lower and upper
    bounds of the phase transition, respectively. The gray band
    corresponds to the phase transition region of the
    EOS. \textit{Center panel:} Same as the left panel, but for models
    $\rm T9_{1.3}^{0}a_{0.4}$ and $\rm
    T9_{1.3}^{-0.01}a_{0.4}$. \textit{Right panel:} Same as the left
    and center panels, but for models $\rm T9_{1.3}^{0}a_{0.6}$ and
    $\rm T9_{1.3}^{-0.01}a_{0.6}$.}
  
  \label{fig:rotating}
\end{figure*}
In the case of rotation we focus on the T9 EOS. We consider models
with the same central energy density as the non-rotating model of
$\epsilon_{\rm c}=\SI{1.3e15}{\g\per\cm\cubed}$ albeit with different
values of the angular momentum. We consider these models both under
equilibrium evolution and negative pressure perturbations. We find
that all rotating models naturally migrate toward the second-family
branch on a dynamical timescale. The rotating models we consider
exhibit strong oscillations in the early stages of evolution which are
eventually damped as the models tend toward their respective
lower-density equilibrium counterparts. As in the non-rotating models
discussed in Sec.~\ref{subsubsec:dep_nonrot}, we find that pressure
depletion in rotating models incites strong oscillations. In
particular, rotating models under pressure depletion exhibit prolonged
oscillations in the rest-mass density when compared to the equilibrium
evolution cases. In the following we present the results of our study
on rotation, categorizing by the initial perturbation.

\subsubsection{Equilibrium evolution}\label{subsubsec:euqi_rot}

In the left, middle, and right panels of Fig.~\ref{fig:rotating} we
show the central rest-mass density for models $\rm
T9_{1.3}^{0}a_{0.2}$, $\rm T9_{1.3}^{0}a_{0.4}$, and $\rm
T9_{1.3}^{0}a_{0.6}$, respectively, using solid lines. In the case of
equilibrium evolution, we find that rotating models exhibit
oscillations early in the evolution which are eventually damped as the
models settle toward configurations with lower central densities,
similar to non-rotating models. The early-time oscillations in
rest-mass density for models $\rm T9_{1.3}^{0}a_{0.2}$, $\rm
T9_{1.3}^{0}a_{0.4}$, and $\rm T9_{1.3}^{0}a_{0.6}$ are significantly
damped (such that the oscillations in $\rho_{\rm c, max}$ are
approximately $1-2\%$) within $t\approx40\tau_{\rm dyn}$,
$t\approx45\tau_{\rm dyn}$, and $t\approx60\tau_{\rm dyn}$,
respectively.

Similar to non-rotating models, the oscillations in rest-mass density
are damped on a dynamical timescale. As with non-rotating models, we
also find that the rest-mass density presented in
Fig.~\ref{fig:rotating} never reaches its pre-bounce value, suggesting
that the models are temporarily bouncing to the additional stable
configurations with central energy densities between the hadronic and
quark phases.

\subsubsection{Pressure depletion}\label{subsubsec:dep_rot}
The evolution of the central rest-mass density for models $\rm
T9_{1.3}^{-0.01}a_{0.2}$, $\rm T9_{1.3}^{-0.01}a_{0.4}$, and $\rm
T9_{1.3}^{-0.01}a_{0.6}$ is depicted using dashed lines in the left,
center, and right panels of Fig.~\ref{fig:rotating}, respectively.
Under pressure depletion with rotation, we observe a qualitatively
similar evolution to non-rotating cases early on.  We observe an
initial contraction in the configuration which corresponds to an
increasing central rest-mass density. The contraction eventually halts
at a maximum and the central region proceeds to bounce between the
hadronic and quark phases. For models $\rm T9_{1.3}^{-0.01}a_{0.2}$ and
$\rm T9_{1.3}^{-0.01}a_{0.6}$ we observe two consecutive strong
bounces early on, as the model returns to the peak rest-mass density
once again after the initial contraction. In the late stages of evolution, the
oscillations are damped. Despite the early-time
strong oscillations for rotating models under pressure depletion,
eventually they all tend to settle near their respective lower-density
counterparts.

We find that rotating models with negative pressure perturbations tend
to undergo strong oscillations of the central region between phases
for a prolonged time.  For models $\rm T9_{1.3}^{-0.01}a_{0.2}$, $\rm
T9_{1.3}^{-0.01}a_{0.4}$ and $\rm T9_{1.3}^{-0.01}a_{0.6}$ we find
strong oscillations (where oscillations in $\rho_{\rm c, max}$ are of
approximately $50\%$ in amplitude) for the first $t\approx 60\tau_{\rm
  dyn}$, $t\approx 50\tau_{\rm dyn}$, and $t\approx80\tau_{\rm dyn}$,
respectively. After the stage of strong bouncing, the
oscillations are damped. Rotating, radially oscillating
compact stars can be potential sources of GWs. We discuss the
prospects of detectability for rotating unstable branch hybrid stars
in detail in Sec.~\ref{subsec:disc_GWs}.

As in the non-spinning cases, we find that the initial increase in
density is large enough for the configurations to reach and exceed the
central density of the counterpart stable twin star, but the solution
does not settle there. For this reason we also investigate the
stability of the stable twin star in App.~\ref{app:stable_hybrids}. We
find that these configurations are dynamically stable (as expected
from the turning point theorem), and they do not exhibit large
oscillations that would lead them transition to the unstable
regime. This suggests, that these stable twin star configurations may
need to be reached in a quasi-static way for them to naturally
form. However, this could be challenging in the case of a
core-collapse supernova or the accretion induced collapse of a white
dwarf or even in the case of a white dwarf--neutron star merger,
because as the central density increases the hadronic branch is
encountered first, and if the density increases further (e.g. due to
compression) and crosses over into the phase transition, the EOS will
soften, causing the star to undergo collapse until the density becomes
high enough for a bounce to take place, and then enter the oscillation
cycles we observed here.

Note that the size of the density perturbation influences the
timescale over which the early-time strong oscillations persist, as
discussed in Sec.~\ref{subsubsec:dep_nonrot}. As such, the interplay
between the strong oscillations incited by pressure depletion and the
natural tendency for rotating models to settle to lower central
densities may depend sensitively on both the angular momentum of the
initial configuration and the size of the initial perturbation. We
leave a more in-depth investigation of the interplay between strong
quasi-periodic oscillations and rotation for future work.

\subsection{Minimum mass instability}\label{subsec:minmass}
\begin{figure}
  \centering
  \includegraphics{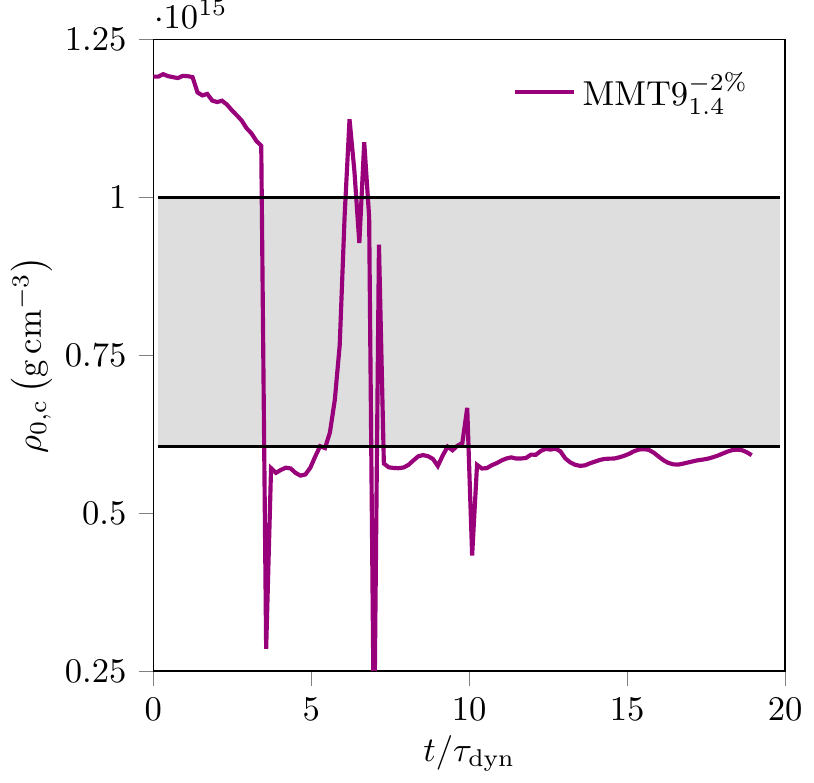}
  \caption{Central rest-mass density $\rho_{0,\rm c}$ for model 
  $\rm MMT9_{1.4}^{-2\%}$ (minimum mass model for the T9 EOS in the case 
  where $2\%$ of the rest mass was removed at the start of the evolution) as a function 
  of time (scaled by the dynamical time $\tau_{\rm dyn}$). The gray band corresponds to
  the phase transition region, and the lower (upper) black line corresponds to the
  start (end) of that region.\\
%  \textit{Right panel:} $L_2$ norm of the Hamiltonian $\mathcal{H}$ 
%  (solid lines) and momentum 
%  $\mathcal{M}$ (dashed lines) constraints for model 
%  $\rm MMT9_{1.4}^{-2\%}$ (magenta lines). Also shown are
%  $\mathcal{H}$ (solid line) and $\mathcal{M}$ (dashed line) 
%  for the case where no mass was removed at the start of the simulation (blue lines).
}
  \label{fig:rhob_minmass}
\end{figure}
\begin{figure*}
  \centering
  \includegraphics{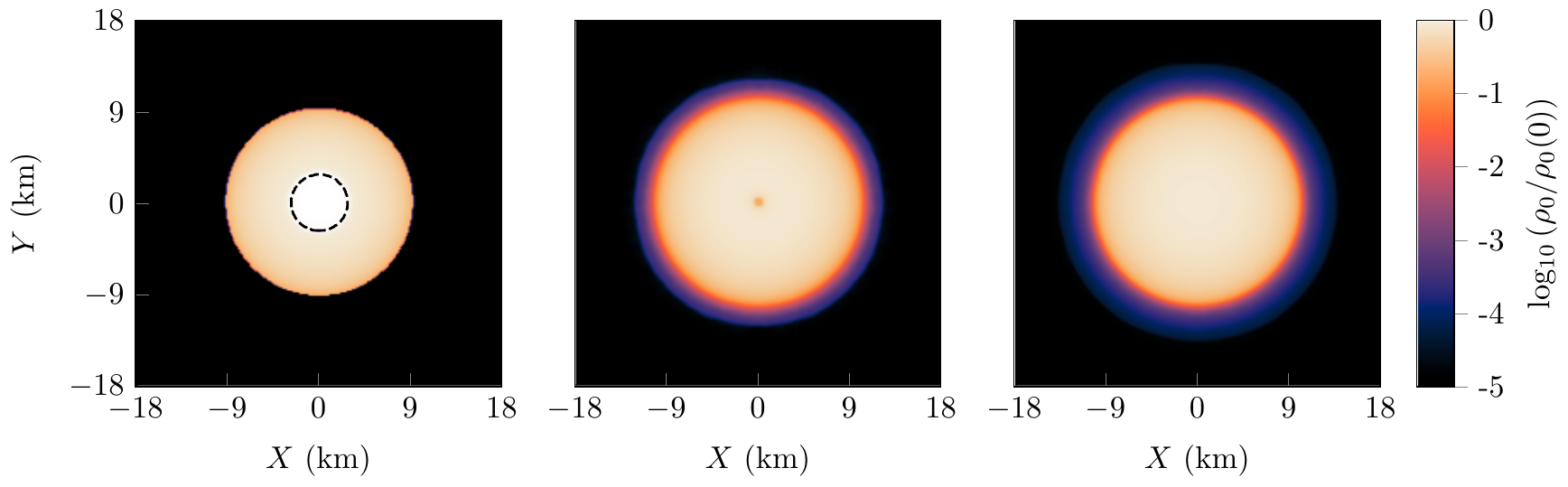}
  \caption{\textit{Left panel:} Equatorial snapshot of the rest-mass
    density for model $\rm MMT9_{1.4}^{-2\%}$ (minimum mass model for
    the T9 EOS in the case where $2\%$ of the rest mass was depleted
    at the start of evolution) at time $t = 0$. The white region
    surrounded by a dashed black line around the stellar center
    corresponds to the quark phase.  \textit{Center panel:} Same
    as the left panel but at time $t\approx 7\tau_{\rm dyn}$. At this
    stage the model has undergone maximal expansion and a small
    under-density can be seen to develop at the center.  \textit{Right
      panel:} Same as the left panel but at time $t\approx 19\tau_{\rm
      dyn}$ near the end of the evolution. By this stage the model has
    roughly settled to a lower-density steady state.  }
  \label{fig:minmass}
\end{figure*}
Neutron stars near the minimum-mass equilibrium configuration, located
on the unstable branch between the stable white-dwarf and stable
neutron-star families, undergo a dynamical instability if the mass is
lowered below the minimum neutron star mass (e.g. if matter is
stripped off the neutron star)~\cite{1977ApJ...215..311C,
  1984SvAL...10..177B, Colpi1989ApJ...339..318C,
  Colpi:1990fe,1998A&A...334..159S}.  We consider this ``minimum-mass
instability'' in the context of minimum-mass hybrid stars, which are
analogously located between the stable neutron star and stable hybrid
star families. The neutron star minimum-mass instability sets in for
configurations in hydrostatic equilibrium wherein the timescale
$\tau_\beta$ associated with the electroweak interactions that
determines chemical equilibrium is comparable to the dynamical
timescale.  In cases where $\tau_\beta/\tau_{\rm dyn} \gg 1$, the
evolution is determined by expansion on a secular timescale, as the
configuration has ample time to oscillate about neighboring
configurations in hydrostatic equilibrium without undergoing beta
decay. On the other hand, if $\tau_\beta/\tau_{\rm dyn} \lesssim 1$,
the configuration undergoes a dynamical instability which ultimately
leads to a spectacular explosion~\cite{Colpi1989ApJ...339..318C,
  Colpi:1990fe,1998A&A...334..159S,2010AstL...36..191M,2020AstL...45..847Y}.
We note that a condition for the base EOSs in this work is chemical
equilibrium between the quark and hadronic
phases~\cite{Alvarez-Castillo2017}, and as such the stellar
configurations we consider are in cold beta-equilibrium.

For minimum-mass neutron stars, the conditions for the instability may
be achieved by mass-stripping, which has the effect of lowering the
total mass while keeping the ratio of electrons to baryons
fixed. Dynamically, a removal of mass similar to that considered
in~\cite{Colpi1989ApJ...339..318C, Colpi:1990fe} in the context of
hybrid stars may take place in close or eccentric black hole-hybrid
star
binaries~\cite{Lattimer:1974slx,1977ApJ...215..311C,East_2012,East:2015yea}. To
incite an initial perturbation similar to that considered
in~\cite{Colpi1989ApJ...339..318C, Colpi:1990fe} for the hybrid stars
considered in our work, we place a rest-mass density cutoff $\rho_{\rm
  0, cutoff}$ on the minimum mass, non-rotating configuration for the
T9 EOS.  To impose the density cutoff, we set all rest-mass densities
below $\rho_{\rm 0, cutoff}$ to the value of the tenuous atmosphere at
the start of the simulation. Motivated by findings which show that
eccentric or close black hole-neutron star binaries can unbind
$\mathcal{O}(10)\%$ or more of the rest mass of one of the binary
components during close
encounters~\cite{East_2012,East:2015yea,PEFS2016}, we set $\rho_{\rm
  0, cutoff} = \SI{1e14}{\g\per\cm\cubed}$, such that approximately
$2\%$ of the initial rest mass is removed. The removal of $2\%$ of the
rest mass also ensures that the constraint violations remain small.

In Fig.~\ref{fig:rhob_minmass} we show the central rest-mass density
for model $\rm MMT9_{1.4}^{-2\%}$ as a function of time. In
Fig.~\ref{fig:minmass} we also show equatorial snapshots of the
rest-mass density at times $t=0$ (left panel), $t\approx 7 t_{\rm
  dyn}$ (center panel) and $t=19t_{\rm dyn}$ (right panel).  We find
an initial expansion of the model as the central rest-mass density
decreases quickly, and falls below the upper density of the phase
transition region, which marks the disappearance of the quark core. At
the peak of the expansion, a small under-density momentarily develops
at the center of the configuration (shown in the center panel of
Fig.~\ref{fig:minmass}). The expansion eventually halts and the core
partially re-collapses. The core oscillates strongly between the quark
and hadronic phases, but eventually settles with a central density
lower than that of the original minimum-mass hybrid solution (as shown
in the right panel of Fig.~\ref{fig:minmass}).  In short, we do not
observe a runaway expansion of the model analogous to the explosions
observed for minimum mass neutron stars. After $t\approx 10 \tau_{\rm
  dyn}$ the model exhibits small ($\sim 5\%$) oscillations in the
central rest mass density, close to the central rest-mass density
corresponding to the hadronic-branch model with rest mass $M_0 \approx
0.98 M_0^{\rm MMT9}$, where $M_0^{\rm MMT9}$ is the rest mass of the
initial minimum-mass hybrid solution. We find that the rest mass of
the initial configuration, after $2\%$ has been removed, is conserved
to within 1 part in $10^{5}$.

The evolution of model $\rm MMT9_{1.4}^{-2\%}$ suggests that there is
enough binding energy present in the initial minimum-mass model to
keep the configuration bound despite the initial expansion. We note
that the results presented in this section are an exploratory
case-study into the minimum-mass instability for hybrid stars.  More
studies are required to definitively state that the minimum-mass
instability does not lead hybrid stars to explode.

\section{Discussion}\label{sec:discussion}
In this section we discuss the results presented in
Sec.~\ref{sec:res}. In particular, we discuss the final state of
rotating models in the context of constant rest mass $M_0$ sequences
and discuss the GW signals associated with the evolution of rotating
models which exhibit strong oscillations.

 \subsection{Final state in the context of evolutionary sequences}
 \begin{figure}
  \centering
  \includegraphics{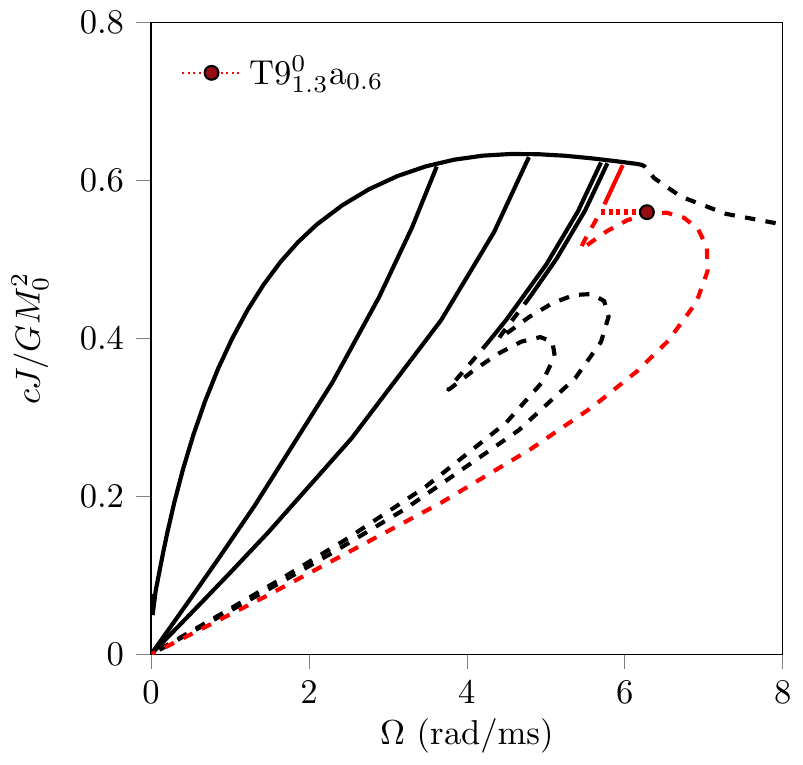}
  \caption{Dimensionless measure of spin, (where $J$ is the total
    angular momentum and $M_0$ is the rest mass) as a function of
    angular velocity for constant rest mass sequences corresponding to
    the T9 EOS. The solid (dashed) lines correspond to sequences that
    only include models with central rest-mass densities below (above)
    $\rho_{0, \rm tr}^i$ as listed in Tab.~\ref{tab:EOS_props}.  The
    red solid and dashed lines correspond to the sequence with rest
    mass fixed to that of the initial configuration of model $\rm
    T9_{1.3}^{0}a_{0.6}$.  We mark model $\rm T9_{1.3}^{0}a_{0.6}$
    with a dark red circle. We also draw a (dotted) line of constant
    angular momentum representing the expected evolution of model $\rm
    T9_{1.3}^{0}a_{0.6}$.}
  \label{fig:Om_J_T9}
\end{figure}
A solution space feature associated with the quasi-radial
turning-point instability is that unstable branch models near the
turning-point may spin up as they lose angular momentum along
sequences of constant rest mass $M_0$ (so called `evolutionary
sequences')~\cite{CST94a, CST94b}.  This type of evolution holds in
situations where the dynamics happen on a long enough timescale that
the models have time to settle into a series of neighboring equilibria
as they evolve. However, in marginally stable or unstable cases where
the timescale associated with the key features of evolution is
shorter, the angular momentum is expected to remain approximately
constant as the models spin up~\cite{Glendenning1997,
  Zdunik:2005kh}. For EOSs with strong phase transitions this
`back-bending' instability is coupled to the dynamical migration of
stars between phases~\cite{Dimmelmeier:2009vw}.  In this section we
discuss the evolution and final states for the set of rotating models
in our study in the context of their respective evolutionary
sequences.

In Fig.~\ref{fig:Om_J_T9} we show the dimensionless angular momentum
as a function of the angular velocity for several evolutionary
sequences corresponding to the T9 EOS. For each sequence, we use solid
(dashed) lines to mark the segments where hadronic (hybrid) stars
reside, such that their central rest-mass densities fall below (above)
$\rho_{0, \rm tr}^i$ as listed in Tab.~\ref{tab:EOS_props}.  We
highlight the evolutionary sequence corresponding to the $\rm
T9_{1.3}^{0}a_{0.6}$ model using red lines and mark the corresponding
initial configuration with a circle of the same color.
In~\cite{Dimmelmeier:2009vw} it was observed that marginally stable
rotating configurations could be forced toward steady states in the
stable hybrid branch by use of density perturbations. It was found
that these configurations tend to spin up while keeping a roughly
constant angular momentum as they settle into the stable hybrid
branch. For the rotating models in our study we find that, even with
pressure depletion (which tends to temporarily drive models away from
the hadronic branch -- the second family), rotating unstable
configurations ultimately settle into the stable hadronic branch.

The tendency for the unstable branch models in our study to migrate
toward the second-family branch happens on a dynamical timescale. The
angular momentum in each of the rotating cases decreases by less than
approximately $0.5\%$, consistent with the absence of angular momentum
emission in gravitational waves from axisymmetric configurations, and
with the level of angular momentum conservation that our code
typically achieves.  An approximation of the angular velocity using
coordinate velocities, i.e., $\Omega=v^\phi=d\phi/dt$, reveals that
our initial configurations tend to spin down as they settle into the
hadronic branch. We note that the such an approximation of the angular
velocity is not gauge invariant, and we only use it as a rough
indicator to discern whether the angular velocity tends to decrease or
increase as the models evolve.  The initial configurations discussed
in~\cite{Dimmelmeier:2009vw} are analogously positioned near the cusp
of the red dashed line in Fig.~\ref{fig:Om_J_T9} and evolve toward the
right on a line of roughly constant angular momentum (for instance,
see Fig.~4 of~\cite{Dimmelmeier:2009vw}). The $\rm
T9_{1.3}^{0}a_{0.6}$ model (represented by the red circle in
Fig.~\ref{fig:Om_J_T9}) instead evolves toward the left roughly along
the dotted line corresponding to constant angular momentum. The
migration of model $\rm T9_{1.3}^{0}a_{0.6}$ toward the stable
hadronic branch happens on a dynamical timescale and we can interpret
the configuration as quickly transitioning from its initial state to
its final state (after a period of oscillation early on) while keeping
$J$ roughly constant, suggesting that the model does not evolve along
an evolutionary sequence. The final state of model $\rm
T9_{1.3}^{0}a_{0.6}$ should roughly reside at the endpoint of the red
dotted line which meets the solid red line (consisting of hadronic
models) in Fig.~\ref{fig:Om_J_T9}.  We note that the angular velocity
of the model marked by the circle in Fig.~\ref{fig:Om_J_T9} 
is much larger than that of the galactic
population of neutron stars~\cite{2006AAS...20720907H,
  2014A&A...566A..64P}. We focus on this model for illustrative
purposes, and find that the same general arguments can be made for the
models in our study with lower angular velocities.

\subsection{Gravitational waves}\label{subsec:disc_GWs}
\begin{figure*}
  \centering
  \includegraphics{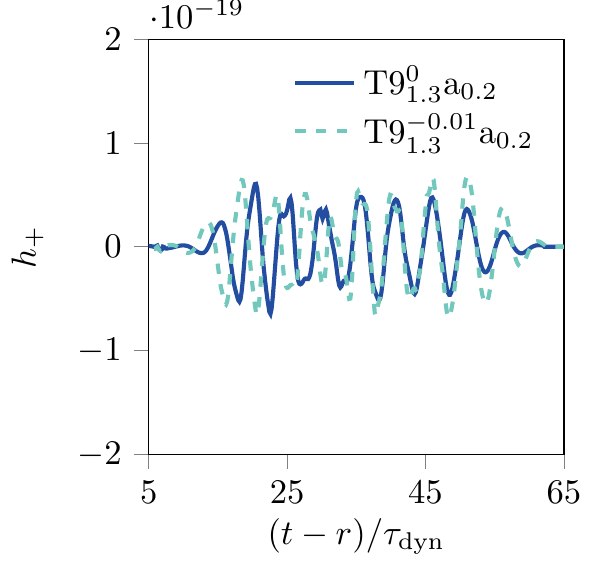}
  \includegraphics{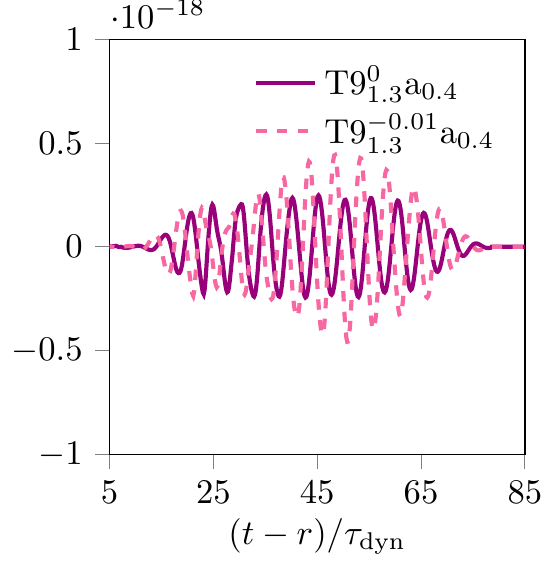}
  \includegraphics{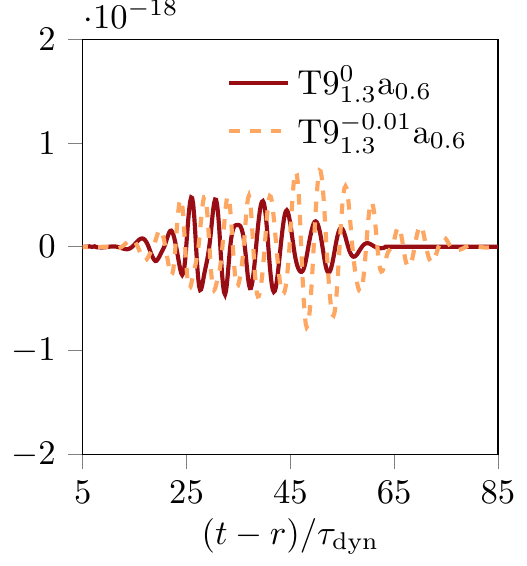}
  \caption{\textit{Left panel:} Plus polarization of the GW strain
    $h_+$ as a function of time (scaled by the dynamical time
    $\tau_{\rm dyn}$) for models $\rm T9_{1.3}^{0}a_{0.2}$ (solid
    line) and $\rm T9_{1.3}^{-0.01}a_{0.2}$ (dashed line) at a
    distance of $10$ kpc assuming an ideal detector orientation.
    \textit{Center panel:} Same as the left panel, but for models $\rm
    T9_{1.3}^{0}a_{0.4}$ and $\rm
    T9_{1.3}^{-0.01}a_{0.4}$. \textit{Right panel:} Same as the left
    and center panels, but for models $\rm T9_{1.3}^{0}a_{0.6}$ and
    $\rm T9_{1.3}^{-0.01}a_{0.6}$.}
  
  \label{fig:GWs}
\end{figure*}

\begin{figure}
  \centering
  \includegraphics{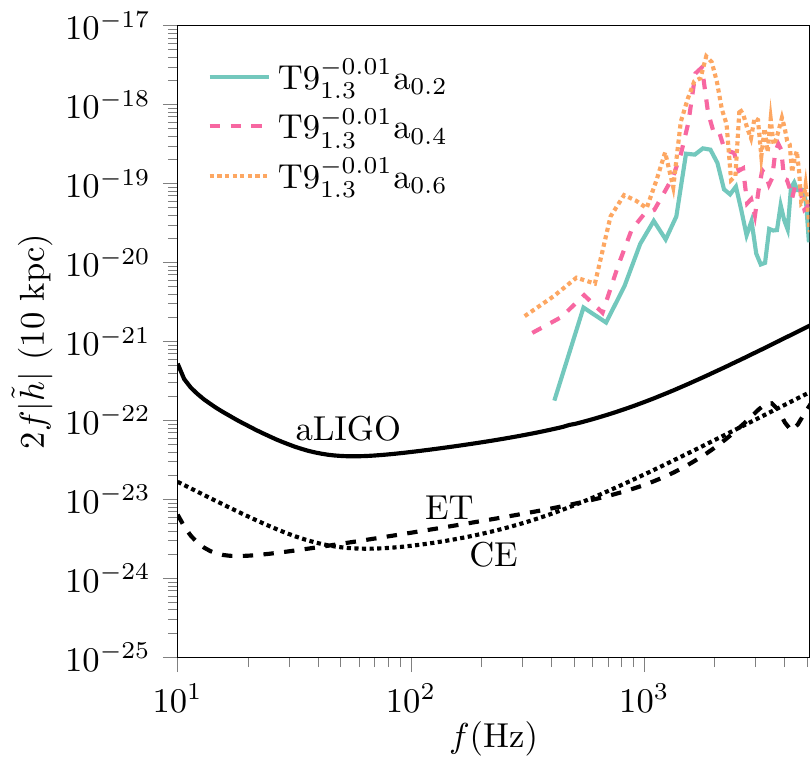}
  \caption{Characteristic strain $h_{\rm
      c}$ for a source at \SI{10}{kpc} for models $\rm
    T9_{1.3}^{-0.01}a_{0.2}$ (blue solid line), $\rm
    T9_{1.3}^{-0.01}a_{0.4}$ (magenta dashed line) and $\rm
    T9_{1.3}^{-0.01}a_{0.6}$ (orange dotted line).  We also show the
    noise curves for future GW observatories including Advanced LIGO
    (upper line, labeled aLIGO) Einstein Telescope (middle line,
    labeled ET), and Cosmic Explorer (lower line, labeled CE).}
  \label{fig:noise_curves}
\end{figure}

We now turn to the possibility that an unstable branch hybrid
configuration could arise as a result of different astrophysical
phenomena~\cite{Takahara:1988yd, Gentile:1993ma, Nakazato:2008su,
  Hempel_2009pe, Yasutake:2009kj,
  Paschalidis:2009zz,Paschalidis:2010dh,Paschalidis:2011ez}. Of
particular interest is the prospect of white-dwarf accretion-induced
collapse or that a WDNS merger results in compression of the neutron
star~\cite{Paschalidis:2009zz,Paschalidis:2010dh,Paschalidis:2011ez,Blaschke:2008na,
  Bejger:2011bq,Drago:2015dea}, which moves part of its core into the
phase transition region where the EOS softens and then the neutron
star continues to shrink. The subsequent development may be the
star dynamically crossing into the unstable twin star regime and undergoing
strong oscillations like the ones discussed in this work. Moreover, it
may be possible for stable twin stars to dynamically transition into
the unstable regime if they are sufficiently close to the minimum mass
twin star and are perturbed. Strongly oscillating rotating stars can
be a promising source of gravitational waves, and the oscillations
driven by the repeated changing of phase in unstable configurations
can lead to a characteristic periodicity in the waves. In this section
we discuss the gravitational wave radiation associated with the
systems we consider as a preliminary consideration of the kinds of
gravitational wave signals we may expect from systems that produce
hybrid stars.

We generally find that the gravitational radiation associated with the
evolution of all rotating models in our study may be detectable only
out to the Andromeda galaxy. Depending on the rotation of the system
in question, we find evidence of features in the GWs which may point
to the evolution of unstable branch hybrid stars as a source.  In
Fig.~\ref{fig:GWs} we show the plus polarization $h_+$ of the GW
strain as a function of time for the rotating models in our study,
assuming a source at \SI{10}{kpc}. The solid (dashed) lines in the
left, center, and right panels of Fig.~\ref{fig:GWs} correspond to
models $\rm T9_{1.3}^{0}a_{0.2}$ ($\rm T9_{1.3}^{-0.01}a_{0.2}$), $\rm
T9_{1.3}^{0}a_{0.4}$ ($\rm T9_{1.3}^{-0.01}a_{0.4}$), and $\rm
T9_{1.3}^{0}a_{0.6}$ ($\rm T9_{1.3}^{-0.01}a_{0.6}$), respectively.
We compute the strain using the FFI method~\cite{Reisswig_2011GWs} and
adopt a low-frequency cutoff $f_0=0.2/\tau_{\rm dyn}$ which is lower
than the peak frequency observed in the power spectrum of all
signals. We focus on the dominant $l=2$, $m=0$ mode and optimal
orientation.  We find that, as expected, models with stronger rotation
produce stronger GWs.  Moreover, models with initial negative pressure
perturbations produce stronger gravitational waves than cases wherein
no perturbation is considered, which is consistent with the stronger
oscillations in the rest-mass density observed in cases with pressure
depletion.

In~\cite{Dimmelmeier:2009vw} it was posited that the dynamical
migration of marginally stable hadronic configurations toward the
stable hybrid branch (labeled in that study as a `mini collapse') led
to a possibly detectable burst of GWs at current and future generation
detectors for an event at a distance of $\SI{10}{kpc}$. In our cases
we do not observe a `mini collapse' scenario (since we don't start
with hadronic configurations), but expect the strength of the GWs to
be comparable to the mini collapse scenario
of~\cite{Dimmelmeier:2009vw} because of the similar size of
oscillations within similar mass stars. Thus, we calculate the strain
associated with sources at the same distance as
\cite{Dimmelmeier:2009vw} for comparison. The strong quasi-periodic
oscillations observed throughout the evolution of all rotating models
allows for rotating systems which produce gravitational waves with a
strong periodicity. The signal associated with these systems builds up
strength at roughly constant oscillation frequency as the central
region continues to bounce between phases, thereby increasing the
detectability of such sources. In Fig.~\ref{fig:noise_curves} we show
the characteristic strain $h_{\rm c}=2f\lvert \tilde{h} \rvert$ (where
$\tilde{h}$ is the frequency-domain signal associated with the GWs in
Fig.~\ref{fig:GWs}) for all rotating models with negative pressure
perturbations at a distance of~\SI{10}{kpc} along with the noise
curves of several future GW observatories. The peak frequency for all
signals is $f_{\rm peak}\approx \SI{2}{kHz}$, which falls within a
low-sensitivity band for these detectors. We note that the short lived
hybrid star remnant observed in~\cite{Gieg:2019yzq} underwent
oscillations which produced GWs with a peak frequency of approximately
$\SI{3}{kHz}$, which is comparable to our findings.  For the distance
considered, the GWs for the rotating cases in our study would be
detectable by Advanced Ligo (aLIGO)~\cite{TheLIGOScientific:2014jea},
Einstein Telescope (ET)~\cite{Maggiore:2019uih} and Cosmic Explorer
(CE)~\cite{Reitze:2019iox}. Specifically, we assume a signal-to-noise
ratio (SNR) detection threshold of 8 at each detector
(see~\cite{Moore:2014lga} for the standard definition of SNR we use
here). At this SNR threshold, all rotating models in our study are
detectable at the three observatories we consider. The detectability
increases for models with stronger rotation. We note that near
monochromatic GWs may also be expected from oscillating neutron stars.
We leave the exploration of how the signals associated with the
oscillations of hybrid stars may be distinguishable from those
associated with oscillating neutron stars to future work.

Following a GW signal associated with the inspiral of a WDNS, a
distinct higher-frequency signal may be observed which could indicate
that the remnant underwent strong oscillations between the hadronic
and quark phases, similar to the rotating models in our study. We
intend to study this scenario in a future work. For sources at a
distance of $d_{\rm source} \gtrsim 1$ Mpc, only the models with high
spin (models with $a\geq 0.4$) produce detectable signals at all
observatories. At this increased distance, models $\rm
T9_{1.3}^{0}a_{0.2}$ and $\rm T9_{1.3}^{-0.01}a_{0.2}$ are not
detectable at aLIGO but are detectable at ET and CE, while all other
rotating models in our study are detectable at the observatories we
consider. Thus, third-generation observatories could potentially see
such events out to the Andromeda galaxy.

\section{Conclusion}
\label{sec:conclusion}

Our simulations suggest that unstable branch hybrid configurations
tend to migrate toward the hadronic branch -- the second family of
compact objects -- on a dynamical timescale. We find that different
types of perturbations drive unstable branch hybrid stars away from
the stable hybrid (third-family) branch or incite models to
temporarily undergo strong oscillations. Specifically, we find that
quasi-radial perturbations induced by positive pressure perturbations
drive the stars toward the hadronic (second-family) branch, and that
pressure depletion temporarily drives the stars away from the stable
hadronic branch. Despite being able to temporarily drive the stars
away from settling into a second family configuration, we were not
able to force any configurations to settle into a stable third-family
model. This is despite the fact that during the evolution the maximum
density reaches and exceeds the counterpart stable twin star central
density.

In select cases we confirmed that the final states reached by models
which tend toward the hadronic branch are consistent with evolutions
of the corresponding stable lower-density equilibrium model with the
same rest mass as that of the initial configuration.  In cases
exhibiting small oscillations near the end of the simulations,
features of the evolution (such as a decaying amplitude of rest-mass
density oscillations) suggest that given enough time, all models will
settle into a configuration resembling their respective stable
lower-density counterparts. We find that rotating models also
naturally tend toward the second-family branch.  In rotating cases
with pressure depletion, we find that the quasi-periodic oscillations
can persist on significant timescales.

After investigating the stability of the stable twin star, we find
that these configurations are stable (as expected from the turning
point theorem), and they do not exhibit large oscillations that would
lead them to transition to the unstable regime. This suggests, that
these stable twin star configurations may need to be reached in a
quasi-static way for them to naturally form. However, this could be
challenging in standard astrophysical scenarios where low mass hybrid
stars may be born (such as systems which produce proto-neutron
stars~\cite{Takahara:1988yd, Gentile:1993ma, Nakazato:2008su,
  Hempel_2009pe, Yasutake:2009kj, Most:2018eaw, Weih_2020}, following
white dwarf--neutron star
mergers~\cite{Paschalidis:2009zz,Paschalidis:2010dh,Paschalidis:2011ez,Blaschke:2008na,
  Bejger:2011bq,Drago:2015dea}. This is because as the central density
increases the hadronic branch is encountered first, and if the density
increases further to cross over into the phase transition, the EOS
softens, and the star may undergo collapse until the density becomes
high enough for a bounce to take place and enter the oscillation
cycles we observed here in the case of unstable twin stars. Our
findings suggest that if the hadron-quark phase transition is over a
large energy density range, it may be difficult to form stable twin
stars in nature. If quark matter deconfinement takes place in compact
star interiors, it appears that stable hybrid stars more massive than
the twin star regime may be preferably formed. This also provides a
alternative formation scenario for stable twin stars, whereby a
massive hybrid star just above the maximum mass second-family
configuration first forms, which subsequently loses mass (e.g.,
through winds) and enters the stable twin star regime.

Our results show that for sufficiently rapidly spinning stars, these
quasi-periodic oscillations the stars may undergo, can produce GW
signatures characteristic of sharp phase transition with a large jump
in energy density (driven by the softening and stiffening of the EOS
over) that can be detectable by ground based observatories as long as
these sources are Galactic, but could be detectable out the Andromeda
galaxy by third-generation observatories.

\begin{figure*}
  \centering
  \includegraphics{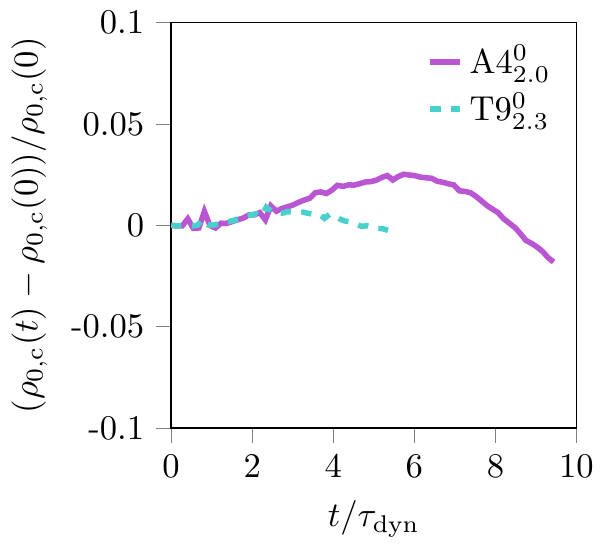}
  \includegraphics{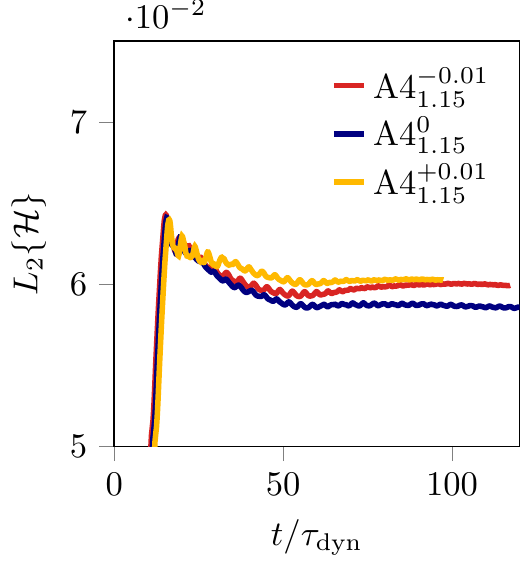}
  \includegraphics{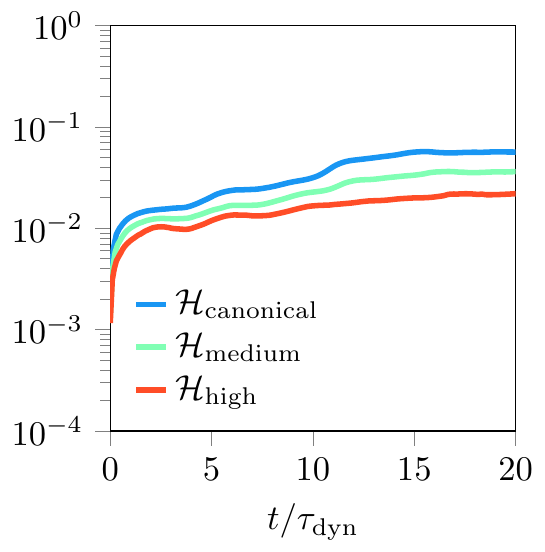}
  \caption{\textit{Left panel}: Fractional change in the central rest
    mass density $\rho_{0, \rm c}$ for stable hybrid branch models
    $\rm A4_{2.0}^{0}$ (magenta solid line) and $\rm T9_{2.3}^{0}$
    (blue dashed line).  \textit{Center panel}: $L_2$ norm of the
    Hamiltonian $\mathcal{H}$ constraint for models $\rm
    A4_{1.15}^{-0.01}$ (red line), $\rm A4_{1.15}^{0}$ (dark blue
    line), and $\rm A4_{1.15}^{+0.01}$ (yellow line) as a function of
    time (scaled by the dynamical time $\tau_{\rm dyn}$). The
    constraints in cases where a pressure perturbation was excited at
    $t=0$ are comparable to the case of equilibrium
    evolution. \textit{Right panel}: $L_2$ norm of the Hamiltonian
    $\mathcal{H}$ constraint as a function of time (scaled by the
    dynamical time $\tau_{\rm dyn}$) for model $\rm A4_{1.25}^{-0.01}$
    at three different grid resolutions. We show results for the
    canonical (blue line), medium (green line) and high (red line)
    resolution grids which employ 64, 80, and 96 grid points per
    hybrid star radius, respectively.}
  \label{fig:stab_test_A4}
\end{figure*}

We conclude by pointing out caveats of the present study.  Our work
provides only some example cases of hybrid hadron quark EOSs that
lead to the emergence of a third-family of compact objects.  Since the
solution space of hybrid stars is
EOS-dependent~\cite{Bozzola:2019tit}, a more extensive study should
treat additional EOSs (with varying ranges over which the
hadron-to-quark phase transition takes place), different types of
perturbations, and probing other parts of the solution space of
equilibrium models. For instance, one may consider a wider variety of
rotating models to understand the interplay between rotation and their
quasi-periodic oscillations. In addition, one can consider the role of
rotation and the growth of non-axisymmetric instabilities
(see~\cite{Paschalidis:2017qmb} for a comprehensive review on the
types of instabilities relevant to rotating relativistic stars).
Finally, differential rotation and magnetic fields can be important
following mergers, which significantly affect the dynamics of possible
BNS merger remnants~\cite{Paschalidis:2015mla, Espino:2019ebx,
  Tsokaros:2019mlz, Ruiz:2019ezy, Nathanail_2020, Ruiz:2020via} or
following a WDNS merger.  We leave the investigation of a more
extensive solution space on the dynamics and mergers to future
studies.

\section{Acknowledgments}
It is a pleasure to thank David Blaschke for useful discussions on
construction methods for matching hadronic and quark phases. We are
grateful to S.\ L.\ Shapiro for access to the code that we used to
build equilibrium models for relativistic hybrid stars, and to
D.\ Alvarez-Castillo, D.\ Blaschke and A.\ Sedrakian for access to the
underlying equations of state. PE and VP acknowledge support from NSF
Grant PHY-1912619. Simulations were in part performed on the
\texttt{Ocelote} and \texttt{ElGato} clusters at the University of
Arizona, the {\tt Comet} cluster at SDSC, and the {\tt Stampede2}
cluster at TACC through XSEDE grant TG-PHY190020. PE is in part
supported by the Marshall Foundation Dissertation Fellowship.

\appendix

\section{Test of stability and resolution study}
\label{app:stability_resolution}

To test that our modified EOSs allow for stable evolutions, we
consider the evolution of a stable branch hybrid star with the A4 EOS
with $\epsilon_c=\SI{2.0e15}{\g\per\cm\cubed}$ (model $\rm
A4_{2.0}^{0}$) and for the T9 EOS with
$\epsilon_c=\SI{2.3e15}{\g\per\cm\cubed}$ (model $\rm T9_{2.3}^{0}$).
In the left panel of Fig.~\ref{fig:stab_test_A4} we show the
fractional change in central rest-mass density for models $\rm
A4_{2.0}^{0}$ and $\rm T9_{2.3}^{0}$.  We note that these models are
not twin stars, and we discuss the stability of stable branch twin
stars in App.~\ref{app:stable_hybrids}.  We find that the test models
are stable over several dynamical timescales, and showed oscillations
in the rest-mass density of at most $2.5\%$ ($1\%$).  These tests
ensure that the migration of unstable branch hybrid stars toward the
hadronic branch are not caused by a lack of stability of hybrid star
models using our modified EOSs.

We also consider the effect of the initial pressure perturbations on
the constraints during evolution. We focus on the cases of model $\rm
A4_{1.15}^{0}$, $\rm A4_{1.15}^{-0.01}$, and $\rm
A4_{1.15}^{+0.01}$. In the center panel of Fig.~\ref{fig:stab_test_A4}
we show the $L_2$ norm of the Hamiltonian $\mathcal{H}$ constraint for
these models.  We find that the constraints are initially small and
quickly saturate to approximately constant values. The size and steady
state value of the Hamiltonian constraint is only weakly affected by
the inclusion of pressure perturbations.  This behavior holds for all
cases where we include initial perturbations, including the case where
we perturb the initial solution with larger or smaller size negative
pressure perturbations (see right panel of Fig.~\ref{fig:dep_nonrot}).
As such, our systematic use of changes to the pressure of form
Eq.~\eqref{eq:p_pert} are justified as small perturbations to the
initial equilibria that are smaller than the truncation error of the
calculation as also illustrated by the convergence plot in the right
panel of Fig.~\ref{fig:stab_test_A4}.

To assess the convergence of our solutions, we perform a resolution
study of the $\rm A4_{1.25}^{-0.01}$ model.  We consider two
additional, higher resolution runs with 1.25 and 1.5 times the
resolution of the canonical grid, which we label as our medium- and
high-resolution runs, respectively. In the right panel of
Fig.~\ref{fig:stab_test_A4} we show the $L_2$ norm of the Hamiltonian
and momentum constraints in the case of the $\rm A4_{1.25}^{-0.01}$
model for our canonical- (blue lines), medium- (green lines), and
high- (red lines) resolution grids, which employ at least 64, 80, and
96 grid-points per hybrid star radius, respectively. We find that the
salient features of evolution (including the initial contraction of
the configuration and subsequent bounce between phases) are generally
invariant through higher resolution simulations. Increasing the
resolution leads to a slightly later initial bounce and a slightly
smaller maximum amplitude of the central rest-mass density, but the
evolution of the model is qualitatively invariant after increasing the
resolution.  We find that the constraints converge to zero with
increased grid resolution at approximately second-order, as expected
for our numerical scheme.

\section{Evolution of stable hybrid twin stars}
\label{app:stable_hybrids}
As an exploration of possible transitions between families of compact
stars, we consider the evolution of stable branch hybrid twin stars.
We focus on the T9 EOS. We construct a model 
which is the stable hybrid twin star counterpart to the third entry on
Tab.~\ref{tab:model_props}. We consider this model under both
equilibrium evolution (which we label $\text{T9}_{1.48}^{0}$) and a
large positive pressure perturbation of $5\%$ in an effort to drive
the model toward a lower density equilibrium (which we label
$\text{T9}_{1.48}^{+0.05}$). In Tab.~\ref{tab:stab_hyb}, we list
some relevant properties of the stable hybrid twin stars we consider.
\begin{table}[htb]\label{tab:stab_hyb}
  \centering
  \caption{Properties of the stable hybrid models we consider, as a
    case study of possible dynamical transitions from stable hybrid
    twin stars to lower density equilibria. For each model, we list
    the model name, the size of the pressure perturbation parameter
    $\xi$ (see Eq.~\eqref{eq:p_pert}), the central energy density
    $\epsilon_{\rm c}$ and rest-mass density $\rho_{0,\rm c}$ (in
    units of $\SI{e15}{\g\per\cm\cubed}$), the gravitational mass $M$
    and rest mass $M_0$ (in units of $M_\odot$), and the compactness
    $C$.  All models in this case study are built using the T9 EOS. }
    \begin{tabular}{l | c c c c c c }\hline  \hline
Model & $\xi$ & $\epsilon_{\rm c}$ & $\rho_{0,\rm c}$ & $M$ & $M_0$ & $C$\\ \hline
$\text{T9}_{1.48}^{0}$      & 0.00 & 1.48 & 1.25 & 1.53 & 1.68 & 0.18 \\ 
$\text{T9}_{1.48}^{+0.05}$ & 0.05 & 1.48 & 1.25 & 1.53 & 1.68 & 0.18 \\ \hline
%$\text{T9}_{1.42}^{0}$      & 0.00 & 1.42 & 1.21 & 1.51 & 1.66 & 0.18 \\ \hline
%$\text{MMT9}_{1.4}^{0}$     & 0.00 & 1.40 & 1.19 & 1.50 & 1.65 & 0.17 \\ \hline
\end{tabular}
 \end{table}
 
\begin{figure}[h]\label{fig:stab_hyb}
  \centering
  \includegraphics{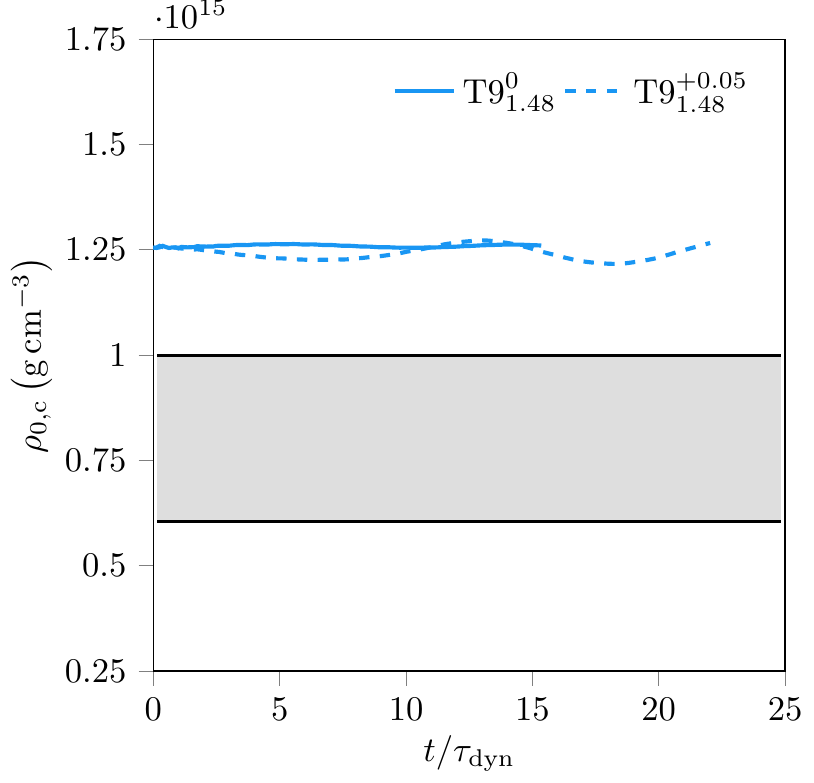}
  \caption{Central rest-mass density $\rho_{0,\rm c}$ as a function of time (scaled by the 
  dynamical time $\tau_{\rm dyn}$) for the models 
  on the stable hybrid branch listed in Tab.~\ref{tab:stab_hyb}. In blue we show results for 
  the evolution of models $\text{T9}_{1.48}^{0}$ and $\text{T9}_{1.48}^{+0.05}$ using solid and
  dashed lines, respectively.\\
}
  \label{fig:stable_hybrid}
\end{figure}

In Fig.~\ref{fig:stab_hyb}, we present the evolution of the maximum
rest-mass density for the models listed in Tab.~\ref{tab:stab_hyb}.
The evolution of these models is both an important test of the
stability of the stable branch hybrid stars built with our modified
EOSs, and serves as a case study of whether stable branch hybrid stars
could conceivably migrate to the unstable regime and thereby
dynamically transition toward the stable hadronic branch. Our findings
suggest that stable branch hybrid stars may not easily transition into
the unstable branch. Model $\text{T9}_{1.48}^{0}$ exhibits very small
radial oscillations over the simulation time of $t\sim 15\tau_{\rm
  dyn}$ (shown using the solid blue line in
Fig.~\ref{fig:stab_hyb}). Model $\text{T9}_{1.48}^{+0.05}$ undergoes
much stronger oscillations than $\text{T9}_{1.48}^{0}$, but the
central rest-mass density never dips below the value of $\rho_{0,\rm
  tr}^f$ for EOS T9 (see Tab.~\ref{tab:EOS_props} for the value of
$\rho_{0,\rm tr}^f$ for this EOS), indicating that the central region
remains in the quark phase and that the configuration is oscillating
about the initial stable hybrid branch model. Oscillations in the
rest-mass density for model $\text{T9}_{1.42}^{+0.05}$ do not grow
above $5\%$.  The evolution of this model demonstrates that it is
stable to radial perturbations, as predicted by the turning-point
theorem. In conclusion, this case study demonstrates that it may
require strong perturbations for configurations on the stable hybrid
twin branch to dynamically transition to the unstable branch and
thereby migrate toward lower density equilibria. On the other hand
approaching these solutions dynamically from the second family
requires strong perturbations whose outcome based on our study of the
unstable twin stars is that the stable twin stars are not the
preferred end states. Instead the second-family solutions appear to be
the natural end states. This suggests that stable twin star solutions
may be reached through quasi-static changes from heavier stable hybrid
stars that lose mass, e.g., through winds, to end up into the twin
star regime. A closer analysis of the dynamics of stable and
marginally stable hybrid twin stars under different types of EOS
descriptions and perturbations is warranted before we can be
conclusively state that stable hybrid twin stars ubiquitously cannot
transition to the unstable regime; we leave a more detailed
investigation of dynamical transitions away from the stable hybrid
branch to future work.

\bibliography{ref}

\end{document}